\documentclass[iop]{emulateapj}
\usepackage{epstopdf}
\usepackage{longtable}
\usepackage{fancyvrb}
\usepackage{multirow}
\usepackage{verbatim}
\usepackage{longtable}
\usepackage[caption=false]{subfig}
\usepackage{float}
\newcommand{\typeI}{Type 1}
\newcommand{\typeII}{Type 2}

\newcommand{\mgii}{\ion{Mg}{2}$\lambda2800$}

\newcommand{\oiii}{[\ion{O}{3}]$\lambda5007$}

\newcommand{\hb}{H$\beta$}

\newcommand{\feii}{\ion{Fe}{2}}

\newcommand{\Hb}{$H$}
\newcommand{\Ib}{$I$}
\newcommand{\Bb}{$B$}
\newcommand{\Vb}{$V$}

\newcommand{\wone}{\emph{W1}}
\newcommand{\wtwo}{\emph{W2}}
\newcommand{\wthree}{\emph{W3}}
\newcommand{\wfour}{\emph{W4}}
\newcommand{\BI}{\Bb-\Ib}
\newcommand{\BV}{\Bb-\Vb}
\newcommand{\sdssu}{$u'$}
\newcommand{\sdssg}{$g'$}
\newcommand{\sdssr}{$r'$}
\newcommand{\sdssi}{$i'$}
\newcommand{\sdssz}{$z'$}
\newcommand{\fuv}{Far-UV}
\newcommand{\nuv}{Near-UV}
\newcommand{\lameff}{$\lambda_{\rm{eff.}}$}

\newcommand{\hst}{\emph{HST}}
\newcommand{\wise}{\emph{WISE}}

\newcommand{\galex}{\emph{GALEX}}
\newcommand{\ch}{\emph{Chandra}}

\newcommand{\galaxyone}{J0841$+$0101}

\newcommand{\galaxytwo}{J0952$+$2552}

\newcommand{\galaxythree}{J1126$+$2944}

\newcommand{\galaxyfour}{J1239$+$5314}

\newcommand{\galaxyfive}{J1322$+$2631}

\newcommand{\galaxysix}{J1356$+$1026}

\newcommand{\reone}{$r_{e,1}$}
\newcommand{\retwo}{$r_{e,2}$}
\newcommand{\recomb}{$r_{e}$}
\newcommand{\prim}{primary}
\newcommand{\second}{secondary}
\newcommand{\n}{$n$}
\newcommand{\none}{$n_{1}$}
\newcommand{\ntwo}{$n_{2}$}
\newcommand{\Arot}{$A_{{\rm rot}}$}

\newcommand{\BIgrad}{$\delta$(\BI)}
\newcommand{\BIavg}{(\BI)$_{\rm{avg.}}$}
\newcommand{\BVgrad}{$\delta$(\BV)}
\newcommand{\BVavg}{(\BV)$_{\rm{avg.}}$}
\newcommand{\BIerrfac}{$5$}

\newcommand{\mone}{$M_{1}$}
\newcommand{\mtwo}{$M_{2}$}

\newcommand{\Mstar}{M$_{\star}$}
\newcommand{\Mlrt}{M$_{\star,\rm{SED}}$}
\newcommand{\Msdss}{M$_{\star,\rm{SDSS}}$}
\newcommand{\SFR}{SFR}
\newcommand{\sSFR}{sSFR}

\newcommand{\SFRlrt}{SFR$_{\rm{SED}}$}
\newcommand{\SFRlrtnorm}{SFR$_{\rm{SED,norm}}$}
\newcommand{\SFRlrtnormdef}{log[\SFRlrt]-log[\SFRctrl]}
\newcommand{\sSFRlrt}{sSFR$_{\rm{SED}}$}
\newcommand{\sSFRlrtnorm}{sSFR$_{\rm{SED,norm}}$}
\newcommand{\sSFRlrtnormdef}{log[\sSFRlrt]-log[\sSFRctrl]}
\newcommand{\SFRsdss}{SFR$_{\rm{SDSS}}$}
\newcommand{\sSFRsdss}{sSFR$_{\rm{SDSS}}$}
\newcommand{\SFRctrl}{$\rm{SFR}_{\rm{ctrl}}$}
\newcommand{\sSFRctrl}{$\rm{sSFR}_{\rm{ctrl}}$}
\newcommand{\Mctrl}{M$_{\star,\rm{ctrl}}$}
\newcommand{\SF}{SF}
\newcommand{\zsdss}{z}
\newcommand{\dbreak}{D4000}
\newcommand{\LUV}{$\nu L_{\nu}$[2800\AA]}
\newcommand{\fMmerg}{$f_{\rm{M,merg.}}$}

\newcommand{\mratio}{\mone/\mtwo}

\newcommand{\loiii}{L$_{\rm{[OIII]}}$}
\newcommand{\LOIIInorm}{L$_{\rm{[OIII],norm}}$}
\newcommand{\LOIIInormdef}{log[\loiii]-log[\loiiictrl]}
\newcommand{\loiiictrl}{L$_{\rm{[OIII],ctrl.}}$}

\newcommand{\SFRu}{M$_{\odot}$ yr$^{-1}$}
\newcommand{\sSFRu}{yr$^{-1}$}
\newcommand{\ageu}{yr}
\newcommand{\physsepu}{kpc}

\newcommand{\gradu}{mag \recomb$^{-1}$}

\newcommand{\ebvu}{mag}
\newcommand{\magu}{mag}
\newcommand{\Mu}{M$_\odot$}
\newcommand{\Lu}{erg s$^{-1}$}

\newcommand{\strlt}{\texttt{STARLIGHT}}
\newcommand{\lrt}{\texttt{LRT}}
\newcommand{\gf}{\textsc{Galfit}}
\newcommand{\se}{\texttt{Source Extractor}}
\newcommand{\el}{\texttt{ellipse}}
\newcommand{\iraf}{IRAF}

\newcommand{\rv}{R$_{V}$}

\newcommand{\ebvagnlrt}{E(B-V)$_{\rm{AGN,SED}}$}

\newcommand{\age}{t$_{\rm{age,SDSS}}$}

\newcommand{\agectrl}{t$_{\rm{age,ctrl}}$}
\newcommand{\metal}{$Z$}

\newcommand{\fitdim}{50}
\newcommand{\ellRefaca}{1}
\newcommand{\ellRefacb}{3}

\newcommand{\mpa}{MPA-JHU}

\newcommand{\ctrlszone}{$18$}

\newcommand{\ctrlsztwo}{$10$}
\newcommand{\galtwoAGNctrlfac}{$7.80$}
\newcommand{\ctrlszthree}{$175$}

\newcommand{\ctrlszfour}{$20$}
\newcommand{\galfourAGNctrlfac}{$2.16$}
\newcommand{\ctrlszfive}{$310$}

\newcommand{\ctrlszsix}{$14$}

\newcommand{\sSFRlrtavg}{$-0.20\pm0.24$}
\newcommand{\sSFRlrtavgoff}{$0.8\sigma$}

\newcommand{\sSFRnormresidfracsig}{$3.7\sigma$}
\newcommand{\sSFRlrtavgCORR}{$-0.57\pm0.16$}
\newcommand{\sSFRlrtavgCORRoff}{$3.5\sigma$}

\newcommand{\sSFRnormresidfracsigCORR}{$3.0\sigma$}

\newcommand{\SFRlrtavgoffscudder}{$1.7\sigma$}

\newcommand{\SFRlrtavgCORRoffscudder}{$5.1\sigma$}

\newcommand{\sSFRnormLOIIInormsig}{$3.7\sigma$}

\newcommand{\sSFRnormLOIIInormsigCORR}{$2.2\sigma$}

\newcommand{\SFRoffset}{$ 36$}
\newcommand{\Moffset}{$  8$}

\newcommand{\parentsz}{ten}
\newcommand{\mergsz}{six}

\newcommand{\delaysz}{five}
\newcommand{\agntypeonesz}{one}
\newcommand{\agntypetwosz}{five}

\newcommand{\galaxyonesdssgemflx}{$0.02$}

\newcommand{\galaxyoneHSTIemflx}{$0.14$}

\newcommand{\meanavgBVreddest}{$0.28$}

\newcommand{\meanavgBVlate}{$0.48$}

\newcommand{\stddeltaBVearlyoff}{$1.87$}
\newcommand{\stdavgBVlateoff}{$1.32$}
\newcommand{\stddeltaBVlateoff}{$1.28$}

\newcommand{\galthreescatuvfrac}{$3.32$}
\newcommand{\galthreescatgfrac}{$0.07$}

\newcommand{\galthreescatIfrac}{$0.01$}

\newcommand{\galthreeAGNBfrac}{$0.99$}

\newcommand{\galfivescatuvfrac}{$35.85$}
\newcommand{\galfivescatgfrac}{$1.31$}

\newcommand{\galfiveAGNHfrac}{$2.85$}

\newcommand{\galsixscatuvfrac}{$15.17$}
\newcommand{\galsixscatgfrac}{$1.40$}

\newcommand{\galsixAGNHfrac}{$0.14$}

\newcommand{\galaxysixexWfour}{$37.32$}
\newcommand{\galtwoscatuvfrac}{$6.25$}
\newcommand{\galtwoscatgfrac}{$1.47$}

\newcommand{\galtwoAGNHfrac}{$41.66$}

\newcommand{\galtwoAGNIfrac}{$31.15$}

\newcommand{\galtwoAGNBfrac}{$63.83$}

\newcommand{\galfourscatuvfrac}{$8.41$}
\newcommand{\galfourscatgfrac}{$0.23$}

\newcommand{\galaxyfourexWthree}{$96.51$}

\newcommand{\galonescatuvfrac}{$88.81$}
\newcommand{\galonescatgfrac}{$8.13$}
\newcommand{\galonescatHfrac}{$0.41$}

\newcommand{\galonescatBfrac}{$4.49$}

\newcommand{\galaxyoneexWthree}{$99.18$}
\newcommand{\galaxyoneexWfour}{$74.59$}

\newcommand{\paperI}{Paper I}

\shorttitle{}
\shortauthors{Barrows et al.}

\begin{document}

\submitted{Accepted for publication in ApJ}

\title{Observational Constraints on Correlated Star Formation and Active Galactic Nuclei in Late-Stage Galaxy Mergers}

\author{R. Scott Barrows$^{1}$, Julia M. Comerford$^{1}$, Nadia L. Zakamska$^{2}$, Michael C. Cooper$^{3}$}
\address{$^{1}$Department of Astrophysical and Planetary Sciences, University of Colorado Boulder, Boulder, CO 80309, USA; Robert.Barrows@Colorado.edu}
\address{$^{2}$Department of Physics and Astronomy, Johns Hopkins University, Bloomberg Center, 3400 N. Charles St., Baltimore, MD 21218, USA}
\address{$^{3}$Center for Galaxy Evolution, Department of Physics and Astronomy, University of California, Irvine, 4129 Frederick Reines Hall, Irvine, CA 92697, USA}

\bibliographystyle{apj}

\begin{abstract}

Galaxy mergers are capable of triggering both star formation and active galactic nuclei (AGN) and therefore may represent an important pathway in the co-evolution of galaxies and supermassive black holes (SMBHs). However, correlated enhancements of merger-induced star formation and AGN triggering may be hidden by the variable conditions and timescales during which they occur. In \paperI, we presented evidence of merger-triggered AGN in a sample of \mergsz~late-stage galaxy mergers ($2-8$ kpc nuclear separations). In this follow-up work, we use multi-wavelength \emph{Hubble Space Telescope} imaging and additional archival data to examine their star-forming properties to test for merger-triggered star formation, and if it is correlated with SMBH growth. We find that the morphological asymmetries are correlated with enhanced specific star formation rates, indicating the presence of merger-triggered star formation. Additionally, the stellar populations become younger with increasing radius from the nucleus, indicating that the merger-induced star formation primarily occurs on global scales. However, we also find that the star formation rate enhancements are consistent with or lower than those of larger separation galaxy pair samples. This result is consistent with simulations predicting a decline of the global star formation rates in late-stage galaxy mergers with $<10$ kpc nuclear separations. Finally, we find that enhancements in specific star formation rate and AGN luminosity are positively correlated, but that an average temporal delay of $\gtrsim 10^{8}$ years likely exists between the peak of global star formation and the onset of AGN triggering in $80\%$ of the systems.

\end{abstract}


\section{Introduction}
\label{sec:intro}

Galaxy mergers can efficiently trigger star formation (\SF) and accretion onto supermassive black holes (SMBHs) that power active galactic nuclei (AGN). While this scenario is likely secondary to that of secular processes for the triggering of most low-luminosity AGN (e.g. in Seyfert galaxies), evidence persists that the most massive SMBHs are grown (in quasar phases) within the most massive galaxies. In this case, a co-evolutionary framework is defined by mergers between massive gas-rich galaxies that are an efficient means of growing both SMBHs and their host galaxy stellar populations. Specifically, merger-induced torques on gas and dust result in both enhanced accretion onto the nuclear SMBHs and enhanced \SF~\citep{Hernquist:1989,Barnes:1991,Mihos:1996,Springel2005,Hopkins2008,Capelo:2016}. This results in a buildup of SMBH mass that tracks the buildup of stellar mass. Indeed, higher luminosity AGN are associated with younger stellar populations \citep{Kauffmann:2003,Wild:2007} and increased \SF~\citep{Heckman:2004,Madau:2014}, while empirical correlations between SMBH masses and stellar bulges do suggest such a proportionality \citep{Gebhardt00,Ferrarese2000,Marconi:Hunt:2003,Gultekin:2009,Bentz:2009c}. The merger pathway also predicts the overall demographics of SMBH host galaxies in which the most massive SMBHs tend to be found in massive elliptical galaxies composed of old stellar populations that are likely the result of gas-rich mergers \citep{Hopkins05,Hopkins2008,Heckman:2014}. 

Individual on-going examples of galaxy-SMBH co-evolution are seen in local ultra-luminous infrared galaxies that contain both rapid \SF~rates (\SFR s) and AGN \citep{Sanders:1988,Sanders:1988a,Canalizo:2001}. Additionally, statistical inferences made with large samples suggest that galaxy mergers drive the growth of the most massive SMBHs and the \SF~in their host galaxies \citep{Rosario:2012}. Other studies have shown that decreasing galaxy pair separations correspond to increases in both the merger fraction of AGN samples \citep{Ellison:2011, Satyapal:2014} and \SFR s in star-forming galaxies \citep{Patton:2011,Scudder:2012,Patton:2013}. However, observationally detecting directly correlated enhancements in \SF~and SMBH accretion among samples of individual galaxy mergers has proven difficult. In particular, while the relevance of galaxy mergers for triggering enhanced \SF~has been robustly established observationally \citep{Joseph:1985,Jogee:2009,Engel:2011,Knapen:2015a,Knapen:2015b}, a corresponding case for AGN triggering is currently tenuous since some existing results favor a connection \citep{Treister:2012,Comerford:Greene:2014,Glikman:2015,Barrows:2017} while others do not \citep{Georgakakis:2009,Kocevski:2012,Simmons:2012,Villforth:2014,Mechtley:2015,Villforth:2016}. 

The lack of a clear connection between \SF~and AGN in galaxy mergers may be due to the different environmental conditions in which they are triggered. In particular, while observations suggest that the AGN merger fraction is preferentially enhanced in major versus minor mergers \citep{Ellison:2011,comerford:2015,Barrows:2017}, a similar dependence on mass ratio is not clear for merger-triggered \SF. Some studies have shown that merger-induced \SF~is strongly negatively correlated with merger mass ratio \citep{Somerville:2001,Cox:2008,Ellison:2008} such that higher \SFR s are seen in major (as opposed to minor) mergers that are more effective at introducing morphological disturbances and randomizing stellar orbits that dynamically evolve galaxies toward early-types \citep{Guo:2016}. On the other hand, minor mergers are theoretically capable of triggering starbursts \citep{Mihos:1994}, and observations show that the \SF~in both major and minor galaxy mergers appears to be similarly enhanced relative to the star-forming main sequence \citep{Willett:2015}. Moreover, a significant fraction of \SF~in the local Universe may occur in galaxies with implicit signs of past minor mergers such as early-type galaxies featuring dust lanes \citep{Shabala:2012} and spiral galaxies with disturbed morphologies \citep{Kaviraj:2014}. Indeed, \citet{Woods:2007} show that the conditions for merger-induced \SF~are more dependent on the strength of the tidal interaction force relative to the galaxy's self gravity rather than merger mass ratio.

Observations of galaxy mergers reveal \SF~on both nuclear scales \citep{Keel:1985,Sanders:1988,Duc:1997} and global scales \citep{Cullen:2006,Elmegreen:2006,Smith:2008,Hancock:2009}. However, the correlation with AGN may be strongest for nuclear \SF~due to a common dependence on the nuclear gas supply. Therefore, the physical extent over which the measured \SF~is integrated can affect the statistical significance of correlations with AGN. For example, one set of simulations finds a statistically significant positive correlation between nuclear ($<100$ pc radii) \SFR~and SMBH accretion rates \citep{Volonteri:2015}. However, when including \SF~on larger physical scales ($<5$ kpc radii), they find a much weaker correlation. When examining the \emph{global} \SF, the only statistically significant connections with AGN appear during late-stage galaxy mergers when the SMBH accretion rates and \SFR s are both large and generate similar luminosities \citep{Rosario:2012,Volonteri:2015}. This prediction is borne out observationally as SMBH accretion only appears correlated with \emph{nuclear} \SF~\citep{Davies:2007,Diamond-Stanic:2012}.

Finally, the level of correspondence between global \SF~and SMBH accretion in galaxy mergers may be affected by the relative chronology of the two. Observationally, temporal delays between the peak of \SFR s and SMBH accretion rates have been identified from AGN residing in host galaxies with relatively old stellar populations. For example, \citet{Schawinski:2009} estimate a time delay of $\sim 100$ Myr between the peak luminosity of X-ray selected AGN and the onset of the decline in \SFR, while \citet{Wild:2010} find that the rise in SMBH accretion rate among optically selected AGN occurs $\sim 250$ Myr after the starburst begins. The reported delays are even longer for radio selected AGN and have been estimated at more than 400 Myr \citep{Shabala:2017} or several galaxy dynamical timescales \citep{Kaviraj:2015b,Kaviraj:2015a}.

Numerical simulations that study the evolution of merger dynamics have produced results that are roughly consistent with these observed temporal delays for \SF~measured over both large (kpc-scale) and small (pc-scale) spatial extents \citep{Hopkins:2012a}. In general, simulations of galaxy mergers find that while the SMBH accretion rate is relatively unaffected until a steep rise after the second pericentric passage that corresponds to nuclear separations of $<10$ kpc \citep{DiMatteo:2005,Springel:2005a,Capelo:2015}, the \SF~experiences a strong peak near the first pericentric passage \citep{Cox:2008,Hopkins2008,Kim:2009,Teyssier:2010,Scudder:2012,Stickley:2014,Renaud:2014}, and on global scales declines gradually throughout the entire subsequent merger process \citep{Capelo:2015}. On the other hand, the nuclear \SF~rises most strongly after the second pericentric passage, accounting for the temporal correspondence seen between merger-triggered \SF~and AGN in simulations \citep{DiMatteo:2005,Springel:2005a,Hopkins2008,Capelo:2015}. Since the stellar populations age after each pericentric passage when the galaxy separations increase \citep{Patton:2011}, the stellar populations are assembled over a much larger fraction of the merger time-frame than the SMBH growth. As a result, the majority of the merger-induced \SF~is completed at nuclear separations of $>30$ kpc \citep{Patton:2013}, thereby producing the measured several hundred Myr ages of the stellar populations once the AGN is observed \citep{Schawinski:2007,Schawinski:2009,Kaviraj:2014,Shabala:2017}. This sequence of stellar and SMBH evolution can account for the fact that \SFR~enhancements in mergers are seen out to galaxy separations of $\sim150$ kpc \citep{Patton:2013} whereas the enhancement of the AGN merger fraction occurs only below $\sim80$ kpc \citep{Ellison:2011,Satyapal:2014}.

Observationally testing these predictions requires a sample of galaxy mergers with sufficient spatial resolution to resolve late galaxy merger stages when AGN are more likely to be triggered, measure morphological disturbances, study the distribution of \SF~at both nuclear and global physical scales, and estimate \SF~histories. In \citet{comerford:2015}, hereafter \paperI, we used a sample of AGN observed with the \emph{Hubble Space Telescope} (\hst) and \ch~to find hints of merger-driven AGN triggering in a sample of late-stage galaxy mergers ($2-8$ kpc nuclear separations). In particular, we found that the frequency with which accretion onto both SMBHs of the progenitor galaxies is triggered tends to be higher for more luminous AGN, in major mergers, and at small nuclear separations. In this work, we search for evidence of merger-driven \SF~and examine how it is related to properties of the host galaxy and the AGN. Ultimately, we place our results within the context of galaxy-SMBH co-evolution. The paper is structured as follows: in Section \ref{sec:sample} we describe the galaxy sample and datasets; in Section \ref{sec:spec_analysis} we describe the spectral analysis; in Section \ref{sec:image_analysis} we describe the image analysis; in Section \ref{sec:results} we describe the results; in Section \ref{sec:discussion} we discuss the results as they pertain to connections between galaxy mergers, \SF, and AGN; and in Section \ref{sec:conclusions} we present our conclusions. Throughout we assume a cosmology defined by $H_{0}=70$ km s$^{-1}$, Mpc$^{-1}$, $\Omega_{M}=0.3$, and $\Omega_{\Lambda}=0.7$.

\section{Sample}
\label{sec:sample}

The sample analyzed in this work consists of \mergsz~galaxy mergers hosting AGN (\agntypeonesz~\typeI~AGN and \agntypetwosz~\typeII~AGN) and that span a redshift (\zsdss) range of \zsdss~$=0.102-0.339$. The AGN classifications are based on the presence of broad emission lines for \typeI~AGN or narrow emission line ratio diagnostics for \typeII~AGN \citep{Baldwin1981,Kewley:2006}. The \typeI~AGN (\galaxytwo) was drawn from the SDSS DR7 Quasar Catalog \citep{Schneider:2010}. Of the \agntypetwosz~\typeII~AGN, four (\galaxyone, \galaxythree, \galaxyfour, and \galaxyfive) were drawn from the \mpa~value-added catalog of physical properties for galaxies and AGN\footnotemark[1]\citep{Kauffmann:2003,Brinchmann:2004}, and the fifth (\galaxysix) was drawn from a catalog of \typeII~quasars \citep{Reyes:2008}. These \mergsz~systems are a subset of a dual AGN candidate sample selected based on double-peaked narrow AGN emission lines (primarily \oiii) in their optical Sloan Digital Sky Survey (SDSS) spectra, a feature that may represent the orbital motion of two AGN narrow line regions (NLRs). From that spectroscopic sample, \parentsz~of those double-peaked systems were selected for follow-up imaging with the \hst~Wide Field Camera 3 in the three filters F160W (\Hb), F814W (\Ib) and F438W (\Bb) and for observations with the \emph{Chandra X-ray Observatory}. 

 \footnotetext[1]{http://www/mpa.mpa-garching.mpg.de/SDSS/}

The analysis in \paperI~used the \hst~imaging to spatially resolve the nuclear regions in search of dual stellar cores and used the \ch~data to spatially constrain the locations of AGN within the systems. \paperI~addressed the interacting nature of these systems, considering a system to be a merger if two galaxy stellar cores are apparent from modeling of the \Hb-band images, finding the above \mergsz~systems to be galaxy mergers. In \paperI, AGN were identified as \ch~detections being spatially coincident with an \oiii~detection and a galaxy stellar core. Merger systems in which only one AGN is spatially identified are referred to as offset AGN and systems in which two AGN are spatially identified are referred to as dual AGN. The analysis in \paperI~found one system that satisfies the dual AGN criteria (\galaxythree) and five that satisfy the offset AGN criteria (\galaxyone, \galaxytwo, \galaxyfour, \galaxyfive, and \galaxysix).

Due to the initial spectroscopic selection and the selection of bright follow-up \ch~targets, these AGN are biased toward high \oiii~luminosities relative to the parent samples from which they were drawn. Based on their integrated \oiii~luminosities from \paperI, the three faintest systems (\galaxythree, \galaxyfour, and \galaxyfive) have \loiii~$=3-5\times 10^{41}$ \Lu~and are comparable to Seyfert galaxies, while the three brightest systems (\galaxyone, \galaxytwo, and \galaxysix) have \loiii~$=2-5\times 10^{42}$ \Lu~and are comparable to quasars \citep{Sanders:1996}.

In this paper, we use the three \hst~filter images for a separate analysis of merger morphologies and galaxy colors. We augment the \hst~imaging with photometric data from the Galaxy Evolution Explorer \citep[\galex;][]{Martin2003}, the SDSS, and the Wide-field Infrared Survey Explorer \citep[\wise;][]{Wright:2010} for spectral energy distribution (SED) modeling and with SDSS optical fiber spectra for synthesized stellar population modeling.

\section{Spectral Analysis}
\label{sec:spec_analysis}

In this section we describe our analysis of the broadband and optical spectra of the \mergsz~merging systems. In Section \ref{sec:sed} we fit galaxy and AGN templates to the broadband photometric SEDs to measure star formation rates and stellar masses. In Section \ref{sec:spec} we fit synthesized stellar templates to the SDSS optical fiber spectra to measure ages of the stellar populations. The physical components included in both the broadband and optical spectral modeling consist of the host galaxy stellar continuum, the AGN continuum, broadened emission lines (in the case of \typeI~AGN), narrow emission lines, and a nuclear obscuring medium. Contribution from the additional components of AGN-heated dust in the mid-IR (MIR), scattered AGN continuum emission in the UV, and line emission from spatially  large NLRs are considered separately. In Section \ref{sec:build_control} we build matched control samples for use in our subsequent analysis.

\subsection{Star Formation Rates and Stellar Masses}
\label{sec:sed}

We build the broadband SEDs using flux densities from three surveys that have observed each galaxy. The highest energy bandpasses are from the \galex~Kron aperture magnitudes in the \fuv~(\lameff~$=1516$~\AA) and \nuv~(\lameff~$=2267$~\AA) detectors. They are followed by the SDSS Data Release 7 (DR7) model magnitudes in the \sdssu~(\lameff~$=3561$~\AA), \sdssg~(\lameff~$=4718$~\AA), \sdssr~(\lameff~$=6185$~\AA), \sdssi~(\lameff~$=7499$~\AA), and \sdssz~(\lameff~$=8961$~\AA) filters. The lowest energy bandpasses are from the \wise~profile-based magnitudes in the \wone~(\lameff~$=3.4$~\micron), \wtwo~(\lameff~$=4.6$~\micron), \wthree~(\lameff~$=12.1$~\micron), and \wfour~(\lameff~$=22.2$~\micron) channels. The choice of these three surveys (up to a combined eleven photometric datapoints covering $\sim0.15$ to $\sim22~\micron$) is to assemble data over most of the $0.03-30~\micron$ range covered by our SED models (described below). All \mergsz~galaxies are detected in all the filters of each survey, with the exception of \galex~for which there are no \fuv~detections of two systems (\galaxythree~and \galaxyfive). In these cases we confirm that the SED model sum does not predict flux greater than the \fuv~sensitivity upper limit at \lameff~$=1516$~\AA. Since the templates do not extend below $0.03\micron$, we do not include \ch~detections in our SED models.

\begin{figure*} $
\begin{array}{cc}
\vspace*{-0.65in} \includegraphics[width=3.5in]{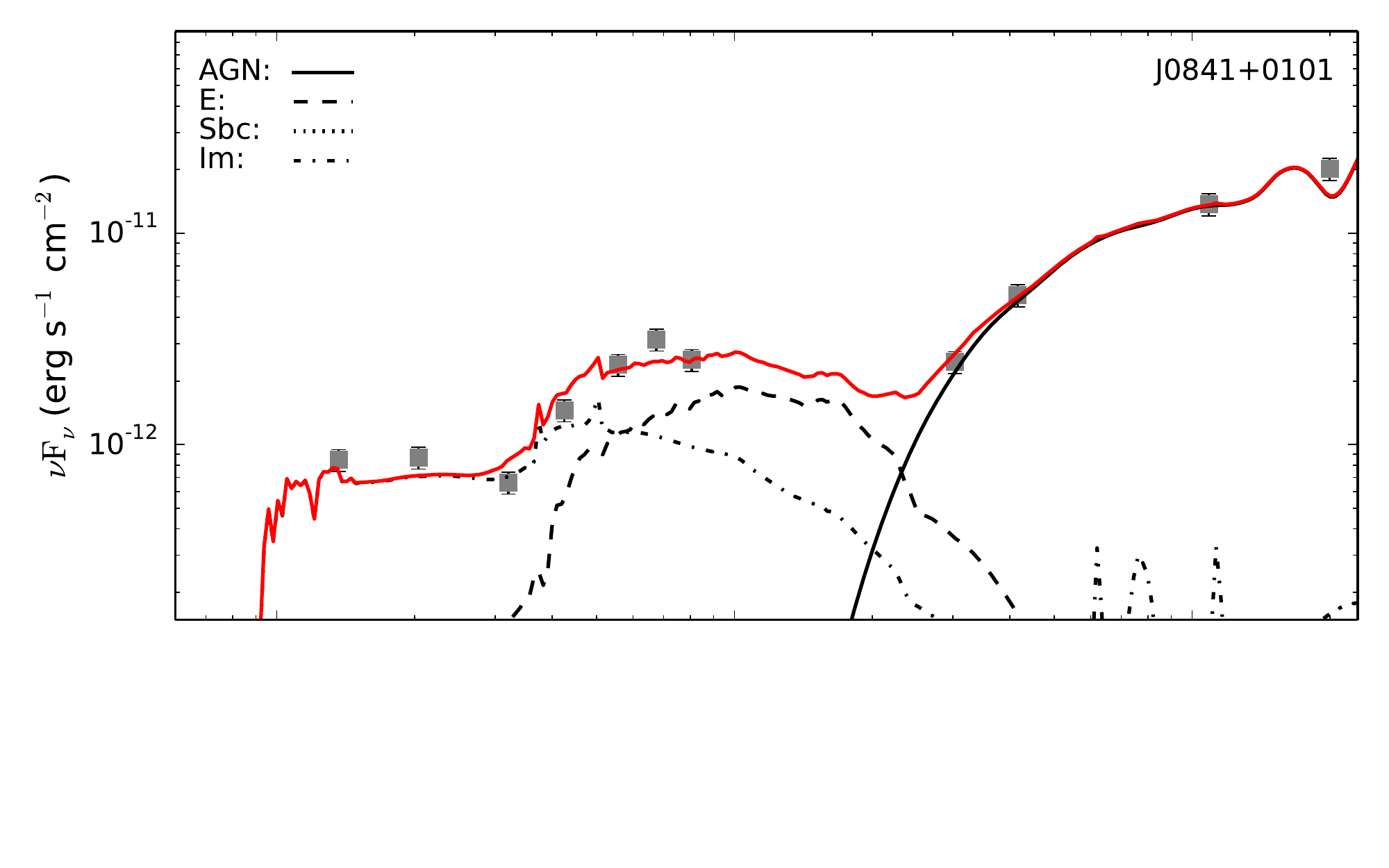} &
\includegraphics[width=3.5in]{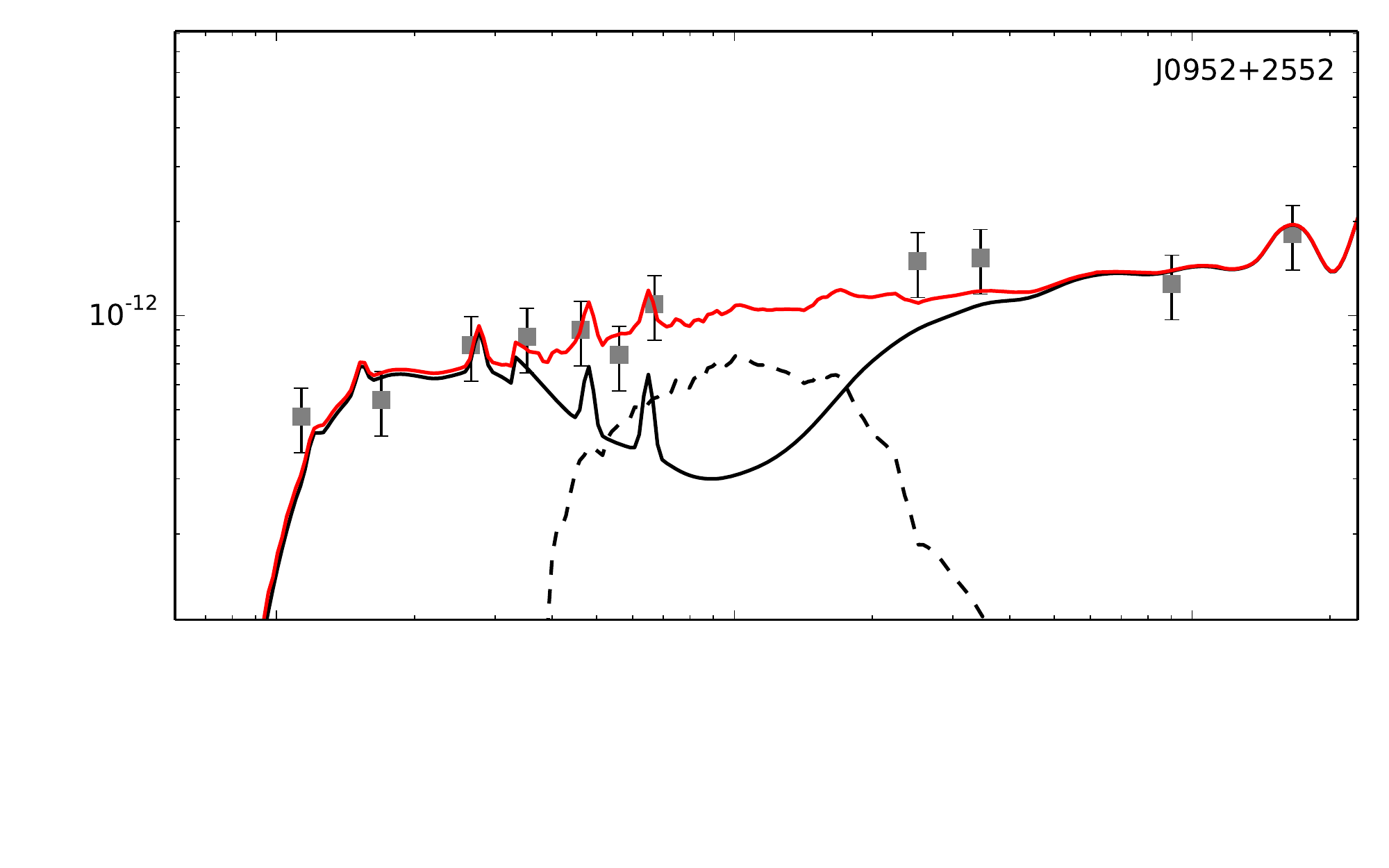} \\
\vspace*{-0.65in} \includegraphics[width=3.5in]{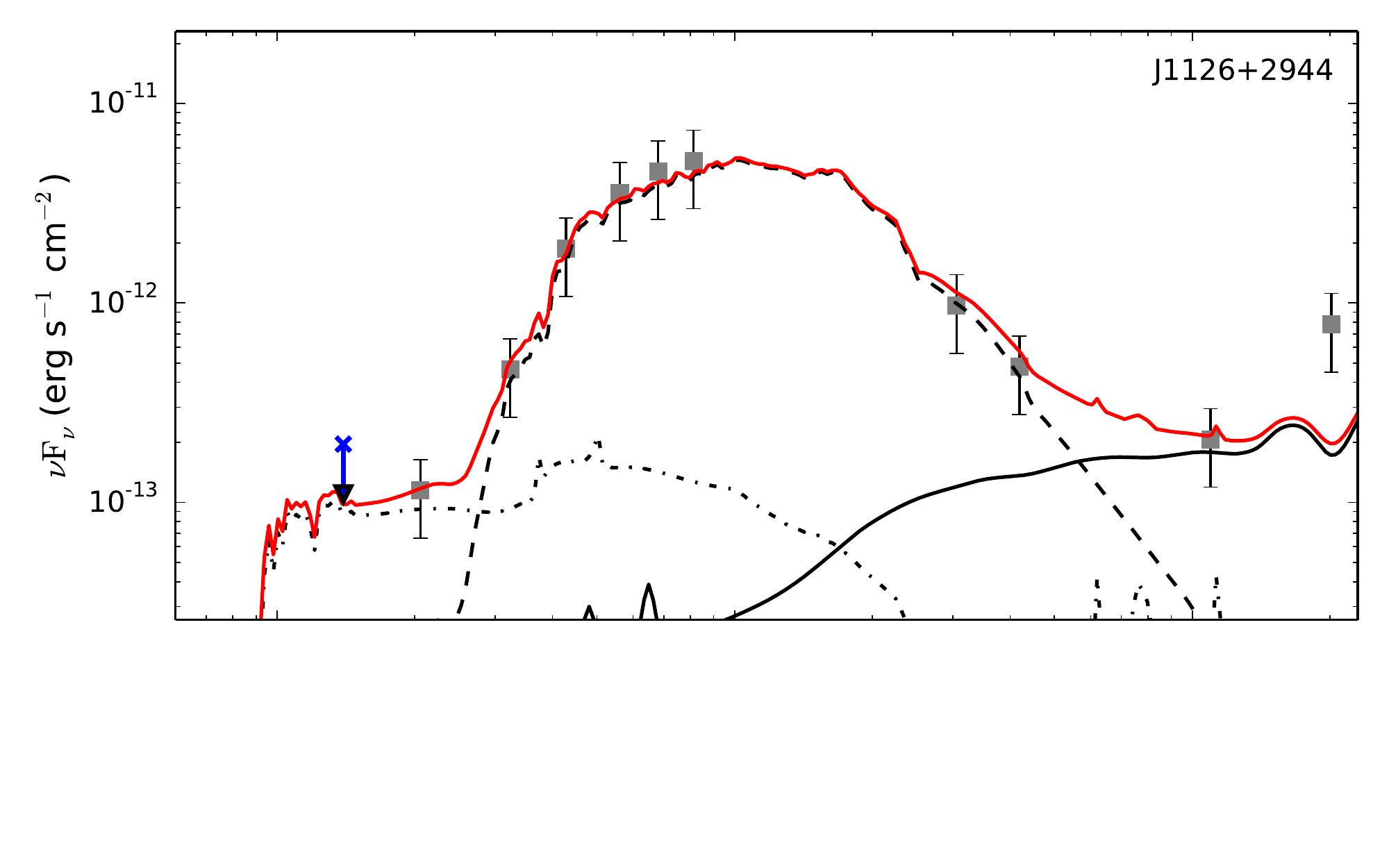} &
\includegraphics[width=3.5in]{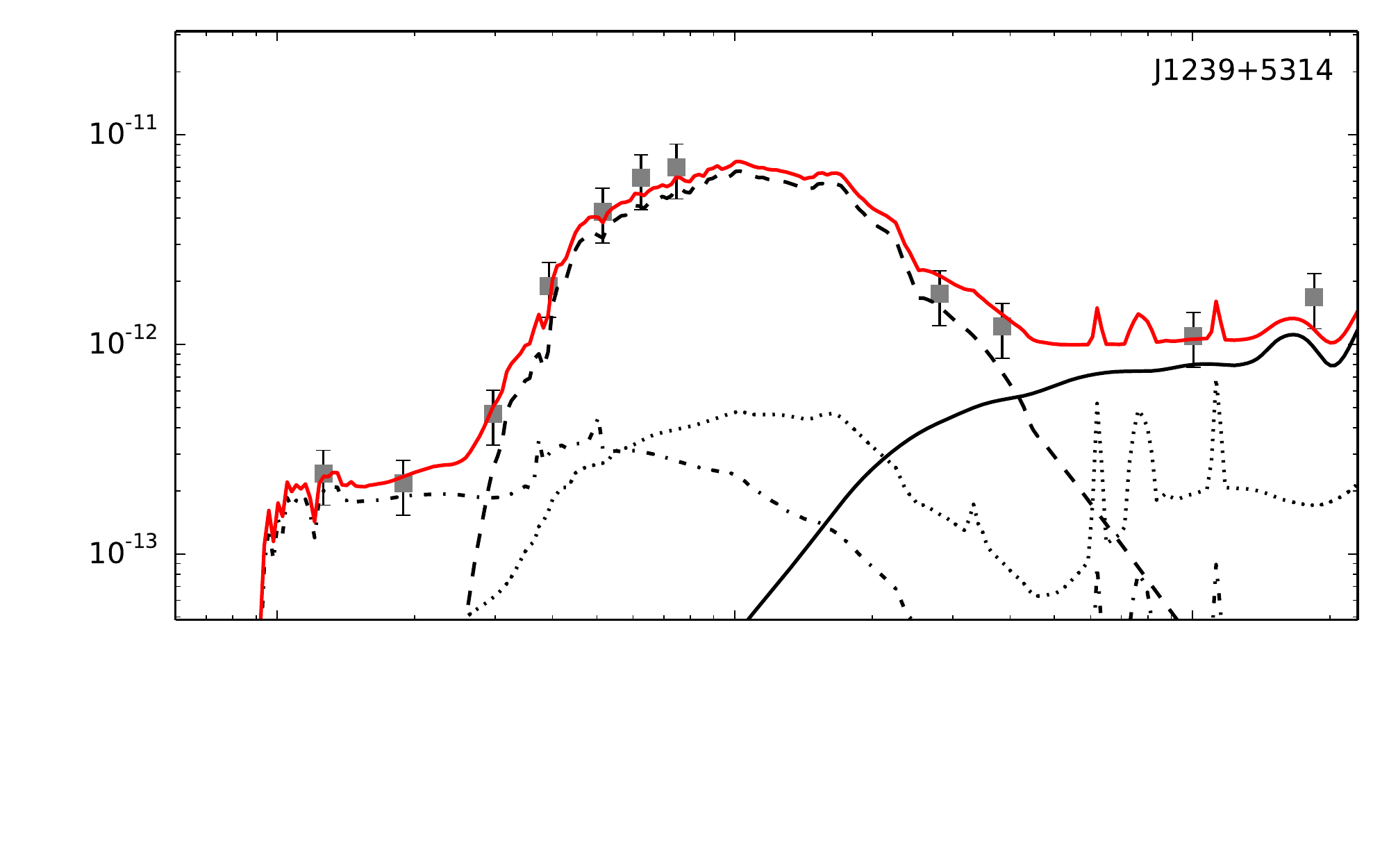} \\
\includegraphics[width=3.5in]{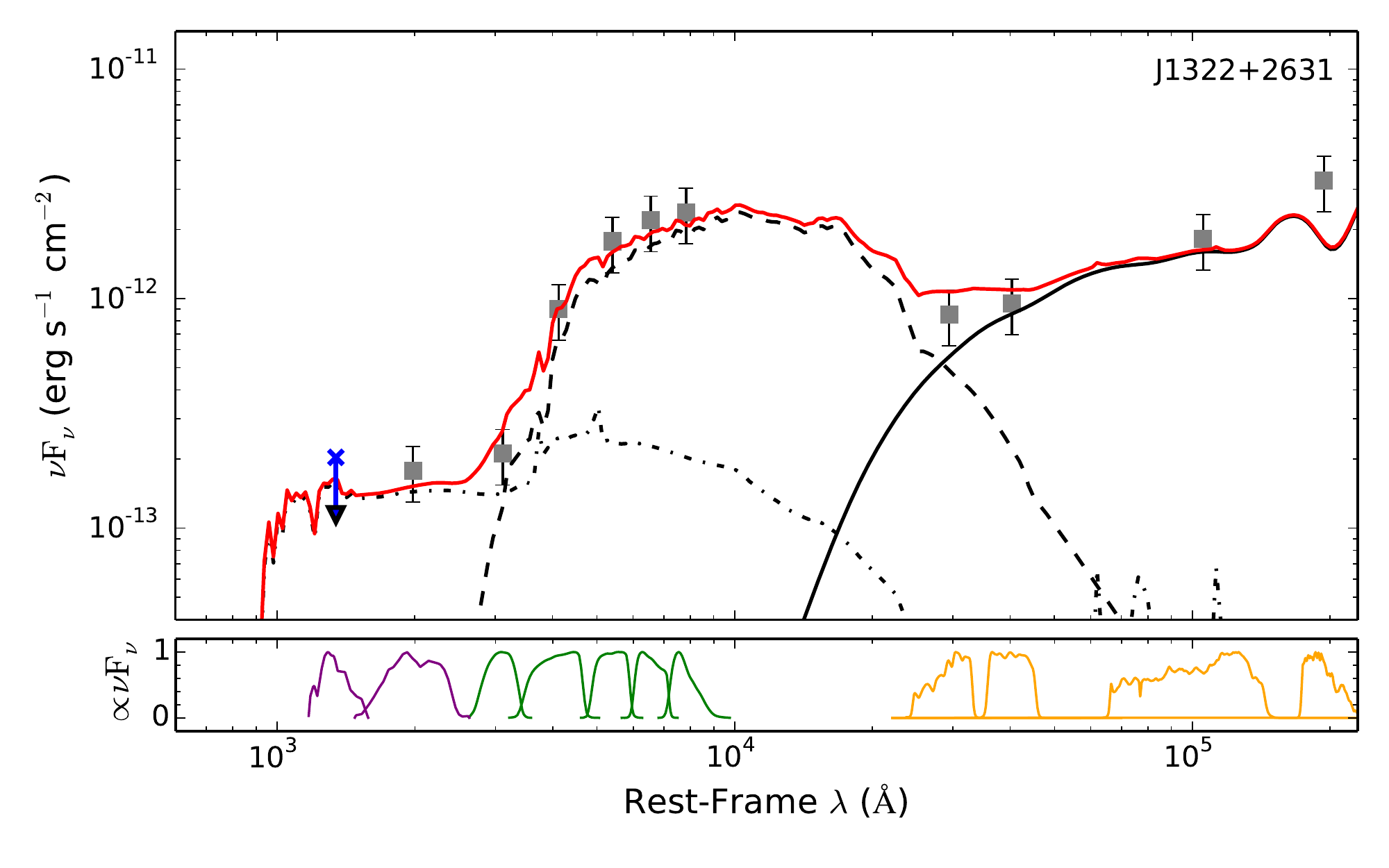} &
\includegraphics[width=3.5in]{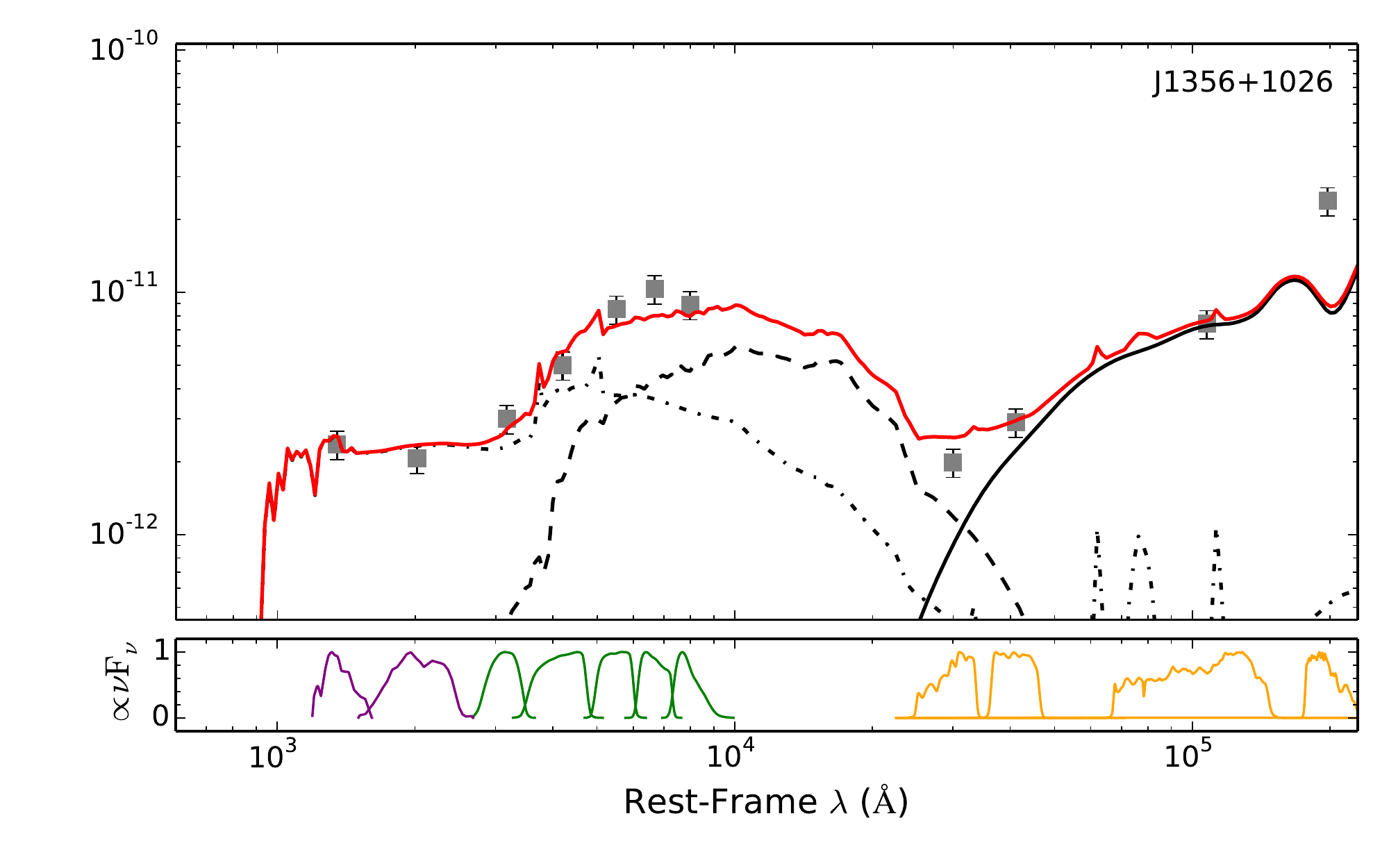}
\end{array} $
\vspace*{-0.05in}
\caption{\footnotesize{Rest-frame SEDs of the \mergsz~merging systems with photometric data points (shown as gray squares) from \galex~(\fuv~and \nuv), SDSS (\sdssu, \sdssg, \sdssr, \sdssi, and \sdssz) and \wise~(\wone, \wtwo, \wthree, and \wfour). The data points are plotted at the filter effective wavelength with their $1\sigma$ upper and lower uncertainties shown. The filter response curves are shown in the bottom of each panel for \galex~(purple), SDSS (green) and \wise~(orange). For each system, the best-fit model sum is shown by the red, solid line. The individual components of the best-fit model are also shown as black lines: AGN with obscuration applied (solid), Elliptical galaxy (dashed), Sbc galaxy (dotted) and Im galaxy (dashed-dotted). For galaxies with no \galex~FUV detection (\galaxythree~and \galaxyfive), the downward arrows indicate the sensitivity upper limit.}}
\label{fig:SED}
\end{figure*}

We model each SED using the Low-Resolution Templates (\lrt) program \citep{Assef2010}. \lrt~models a broadband SED as a linear combination of galaxy templates (elliptical, Sbc Spiral, and Irregular) plus a \typeI~AGN template. The AGN template was developed by \citet{Assef2010} as a combination of power laws plus broad emission lines and is designed to resemble the composite \typeI~quasar template from \citet{Richards2006}. A wavelength-dependent extinction law constructed from the \citet{Cardelli:1989} optical-IR function (for $\lambda>3300$\AA) and the \citet{Gordon:Clayton:1998} UV function (for $\lambda<3300$\AA) is also applied to the AGN template. This extinction term, \ebvagnlrt, effectively accounts for nuclear obscuration and thereby suppresses both the continuum and narrow/broad emission lines of the AGN template. With sufficient extinction, the optical and UV portion of the final SED will resemble that of a \typeII~AGN. We note that, in the case of the dual AGN system (\galaxythree), the two AGN likely experience different levels of nuclear obscuration. However, the two AGN components and their extinction terms are strongly degenerate and we do not have the spectral resolution to deblend them. Therefore, we only include the single extinguished AGN component and note that the nuclear obscuring solution likely reflects a value intermediate to that of the two AGN. The SEDs and best fit models are shown in Figure \ref{fig:SED}.

We sum the best-fit elliptical, Sbc Spiral, and Irregular galaxy templates to measure the rest-frame monochromatic UV ($2800$~\AA) luminosity (\LUV) of each galaxy assuming the SDSS spectroscopic redshift and the cosmology stated in Section \ref{sec:intro}. Then assuming a Salpeter initial mass function we use \LUV~and the UV-based relation of \citet{Madau:1998} to estimate star formation rates (\SFRlrt). We also use the summed galaxy flux to measure the host galaxy stellar masses (\Mlrt) using the \sdssg$-$\sdssr~color, M/L$_{g}$ relation of \citet{Bell:2003}, and the \sdssg- and \sdssr-band $k$-corrections from our SED models. We then combine \SFRlrt~with \Mlrt~to estimate specific star formation rates (\sSFRlrt~$=$~\SFRlrt$/$\Mlrt). In Table \ref{tab:SFR} we list \ebvagnlrt, \SFRlrt, \Mlrt, and \sSFRlrt.

\subsubsection{Additional Sources of Emission}
\label{sec:additional}

In this section we consider additional sources of emission that are not explicitly included in the SED models but may contribute to the broadband photometry and affect our estimates of \SFRlrt~and \Mlrt. In particular, we consider scattered AGN continuum flux, emission line flux from extended NLRs, and flux from AGN-heated dust.

\emph{Scattered light:} While the nuclear continuum source and broad line region are obscured from direct view in \typeII~AGN, a fraction of that light can be scattered into the line-of-sight \citep{Antonucci:1985,Kishimoto:1999}. The flux due to electron and dust scattering rises toward and peaks in the rest-frame UV \citep{Kishimoto:2001,Draine:2003}, and in powerful \typeII~quasars the scattering regions can reach several kpc in size \citep{Zakamska:2005}. As a result, the UV contribution to the SED from scattered flux at large radii may be significant for the more powerful AGN in our sample. The degree of scattering is parameterized by the scattering efficiency: the ratio of scattered flux to that of the intrinsic source \citep{Zakamska:2006}. Since scattered flux in AGN hosts at UV wavelengths is a strong function of the intrinsic AGN luminosity, we estimate the scattered flux at $3000$~\AA~assuming a typical scattering efficiency of $3\%$ \citep{Obied:2016}. 

To quantify the effect of scattered light on our measurements of \SFRlrt~and \Mlrt, we extrapolate the scattered flux to all wavelengths assuming a power-law function ($F_{\lambda}\sim \lambda^{-\alpha}$) with $\alpha=1.5$ \citep{Berk:2001}. Relative to the overall SED flux at $2800$~\AA, we find scattered light fractions of \galonescatuvfrac$\%$, \galtwoscatuvfrac$\%$, \galthreescatuvfrac$\%$, \galfourscatuvfrac$\%$, \galfivescatuvfrac$\%$, and \galsixscatuvfrac$\%$~for \galaxyone, \galaxytwo, \galaxythree, \galaxyfour, \galaxyfive, and \galaxysix. These values are qualitatively similar to the lower limits found for the sample of more luminous \typeII~quasars from \citet{Zakamska:2006} in the F550W filter. Relative to the overall flux in the \sdssg-band, we find significantly lower scattered light fractions of \galonescatgfrac$\%$, \galtwoscatgfrac$\%$, \galthreescatgfrac$\%$, \galfourscatgfrac$\%$, \galfivescatgfrac$\%$, and \galsixscatgfrac$\%$ for \galaxyone, \galaxytwo, \galaxythree, \galaxyfour, \galaxyfive, and \galaxysix. Since the scattered light affects the $2800$~\AA~measurements by factors of $4-40$ more than the \sdssg-band measurements, the \SFRlrt~values are likely significantly more over-estimated than those of \Mlrt, contributing to over-estimates of \sSFRlrt.

\emph{AGN-heated dust:} A dusty obscuring medium is expected to reprocess AGN continuum emission, thereby contributing to excess MIR flux not accounted for by the \typeI~AGN template. Indeed, we see that the SED model significantly under-predicts the \wfour~photometry in some of the systems by \galaxysixexWfour$-$\galaxyoneexWfour$\%$. Three of the four systems with the most under-predicted \wfour~fluxes correspond to the three highest values of \ebvagnlrt, while the system with the least under-predicted \wfour~flux corresponds to the lowest value of \ebvagnlrt. Additionally, after removing \galaxythree~(which hosts two AGN whose separate nuclear obscuring components are not accounted for; Section \ref{sec:sed}), the values of \ebvagnlrt~are inversely correlated with the \wfour~model under-predictions with a Spearman rank statistic of $r=0.70$ and null-hypothesis probability of $p=0.18$.  This connection between excess MIR flux and nuclear extinction is consistent with AGN continuum emission that is absorbed and reradiated by the nuclear obscuring medium. Furthermore, this signature of relatively steeply rising MIR flux is also seen in \citet{Wylezalek:2016}, who find that simple models of dust distributions leave significant excess flux at rest-frame wavelengths of $>10~\micron$ and conclude that it is primarily due to dust heated by the central AGN. 

However, this effect is negligible in the \wthree~band (\galaxyfourexWthree$-$\galaxyoneexWthree) and at similar levels in all higher energy bands. Therefore, we assume that the contribution from AGN heated dust does not significantly affect the modeled fluxes at higher energies and hence the modeled AGN component normalization (used for our scattered flux estimates) is accurate.

\emph{Emission lines:} NLRs are expected to extend beyond the nuclear obscuring medium and therefore suffer less attenuation than the broad emission lines. However, suppression of the AGN template in the SED modeling also suppresses the modeled NLR contribution without accounting extended NLR emission that is not subject to obscuration. Furthermore, the SED modeling does not account for NLRs of above average luminosity (e.g. from powerful AGN or enhanced by AGN outflows/shocks). Therefore, to estimate the actual flux contributions of each narrow emission line to the photometry, we separately convolved the galaxy continuum and narrow emission line components that are accessible in the SDSS optical spectra with the filters used in the SED modeling to produce synthetic magnitudes for each. The ratio of narrow emission line flux to total galaxy flux in each filter represents the maximum contribution of the narrow emission lines to each filter. 

Emission lines from extended or powerful NLRs may be contributing to excess flux and causing the SED to over-estimate the galaxy components. The emission line contributions to the \sdssg-band due to the \oiii$+$\hb~complex are $<$\galaxyonesdssgemflx$\%$. While we do not have spectral access to rest-frame UV emission lines and can not estimate their direct contributions to \LUV, they are likely negligible since the \mgii~line is confined to the broad line region. The overall emission line contribution may be stronger than our estimates since the \mpa~catalog values are normalized to the fiber magnitudes (i.e. they do not include an aperture correction) so that the emission line contributions estimated from the optical spectra may be under-estimates for the whole systems if extended NLRs are present. For example, \galaxysix~is known to contain a large-scale AGN-driven outflow of ionized gas in \hst~imaging that extends beyond the SDSS fiber \citep{Greene:2014}. However, since our estimates of emission line contributions within the fiber are negligible, any further contribution from extended emission is likely to be negligible as well. Therefore, the emission line contributions are smaller than those of scattering in both the \sdssg-band and at $2800$~\AA, leading to the result that the effects of scattering are likely to dominate over those of emission lines.

\begin{figure}
\hspace*{-0.1in} \includegraphics[width=3.5in]{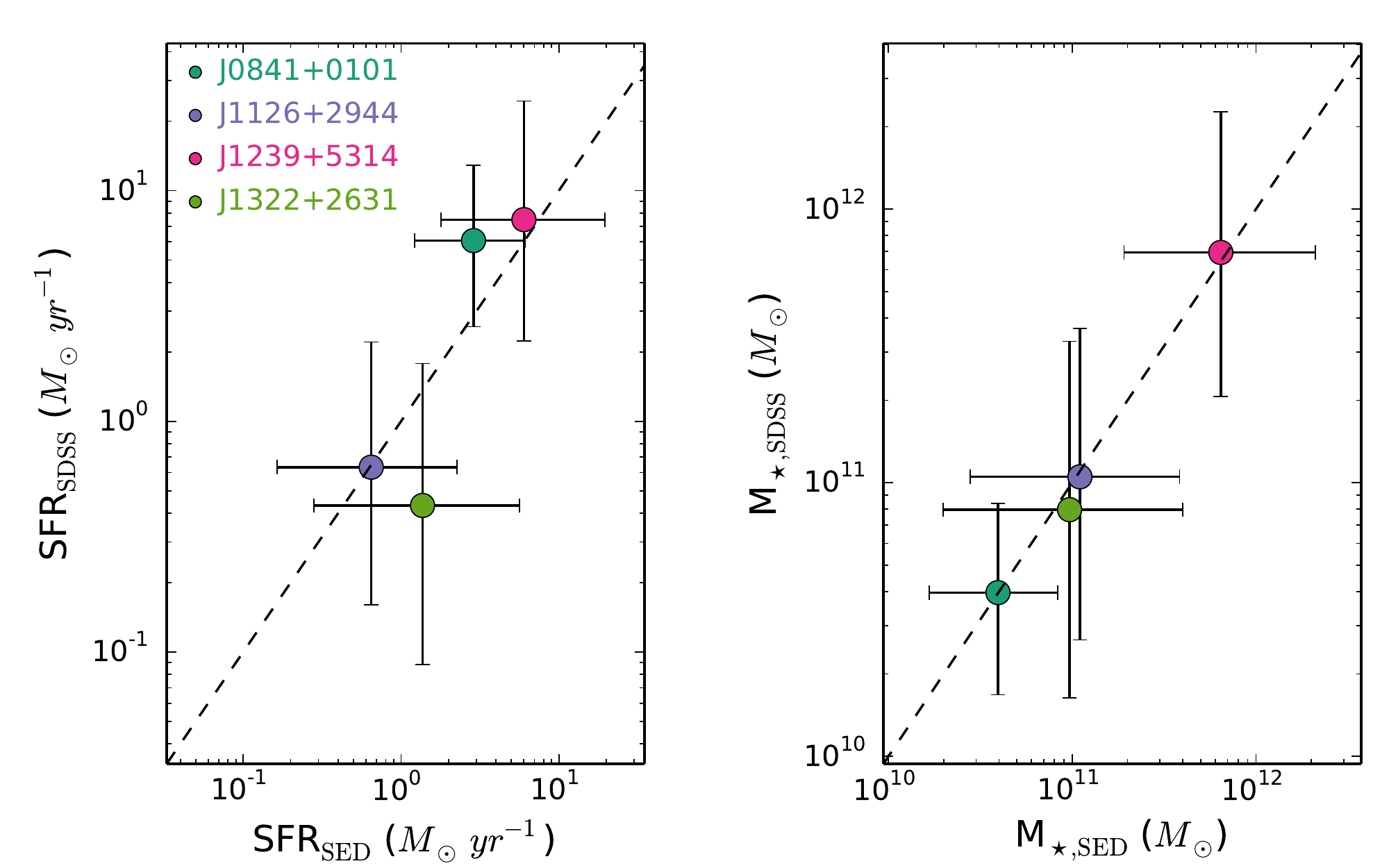}
\caption{\footnotesize{Left: \SFR~estimates from SED modeling versus from the SDSS. Right: \Mstar~estimates from SED modeling versus from the SDSS. In each panel the dashed line represents the one-to-one relation. Note that the values from SED modeling and from the SDSS are consistent within $1\sigma$ among values of \SFR~and \Mstar, though the scatter is larger for the former.}}
\label{fig:plot_lrt_sdss}
\end{figure}

\subsubsection{Comparison Measurements}

For comparison, we also obtained measurements made with the SDSS optical fiber spectra (\Msdss~and \SFRsdss) for the four galaxies in our sample in the \mpa~catalog. Stellar masses in the \mpa~catalog (\Msdss) are generated by fitting a grid of models from \citet{Bruzual:Charlot:2003} to the broad-band SDSS photometry. For AGN, measurements of \sSFRsdss~are based on the $4000$~\AA~break (\dbreak) and the empirical correlation between \dbreak~and \sSFR~calibrated from star-forming galaxies in \citet{Brinchmann:2004}. We compute values of \SFRsdss~as \SFRsdss~$=$~\sSFRsdss~$\times$~\Msdss. 

\begin{figure*}[t!] $
\begin{array}{cc}
\vspace*{-0.3in} \includegraphics[width=3.5in]{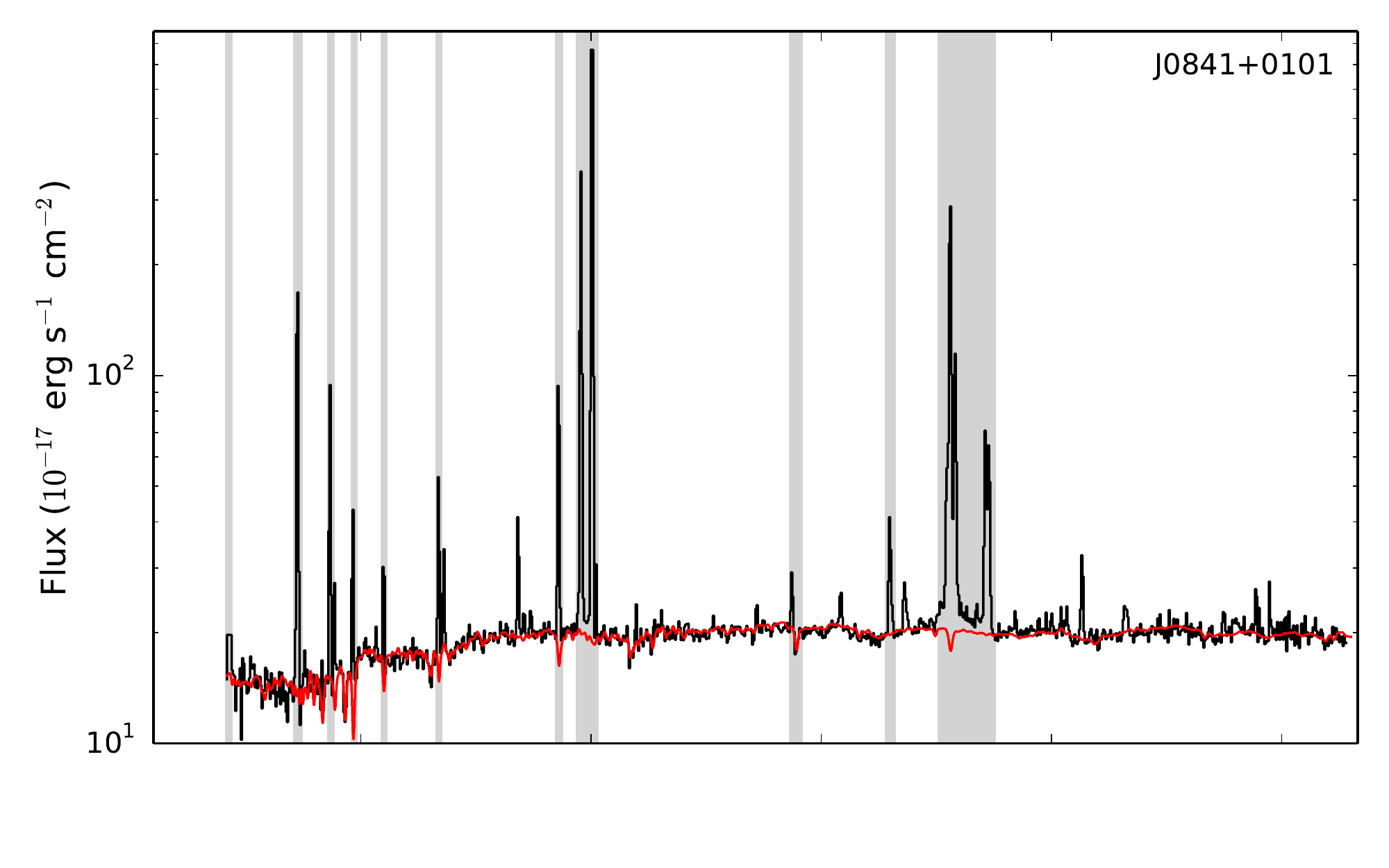} &
\includegraphics[width=3.5in]{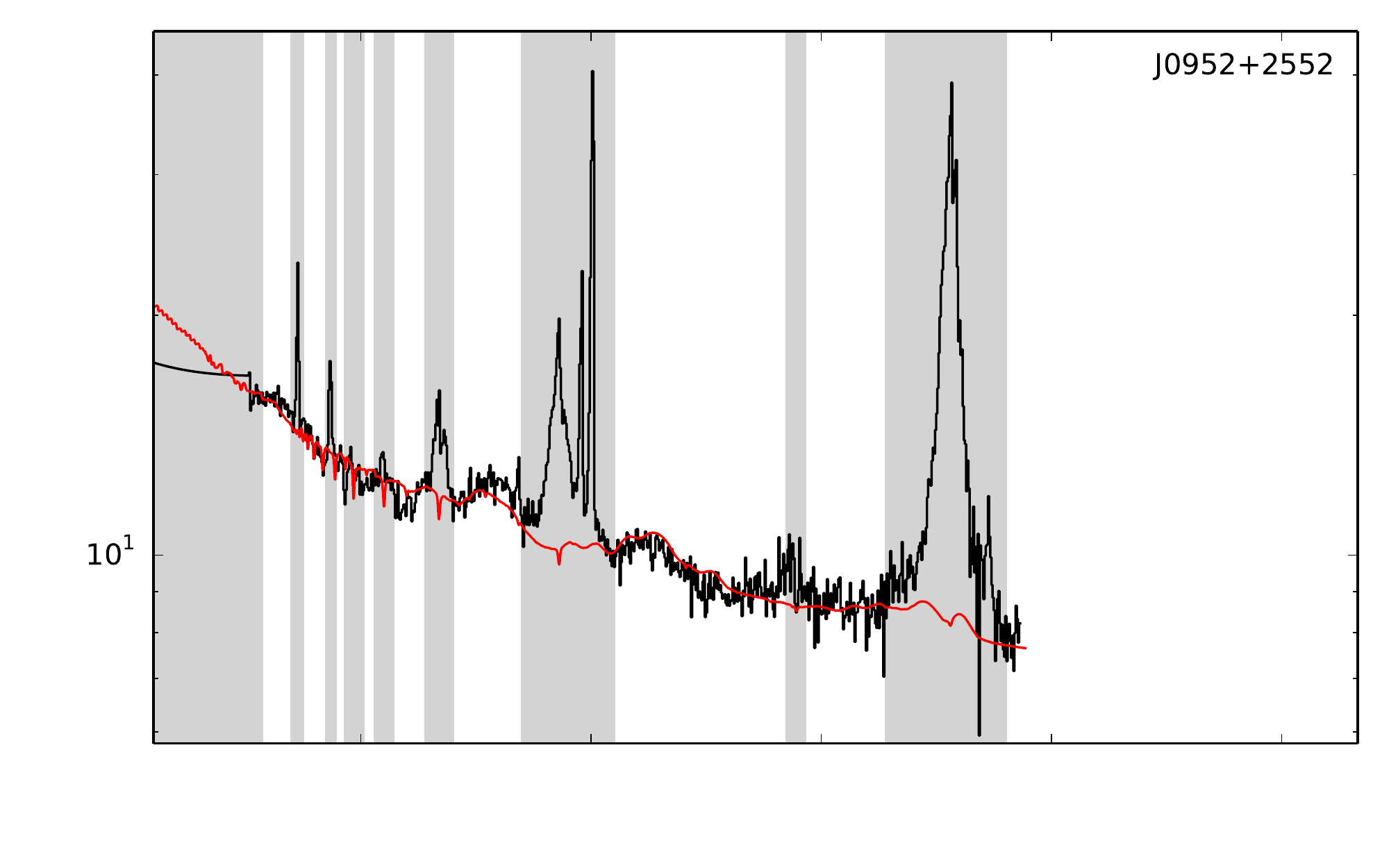} \\
\vspace*{-0.3in} \includegraphics[width=3.5in]{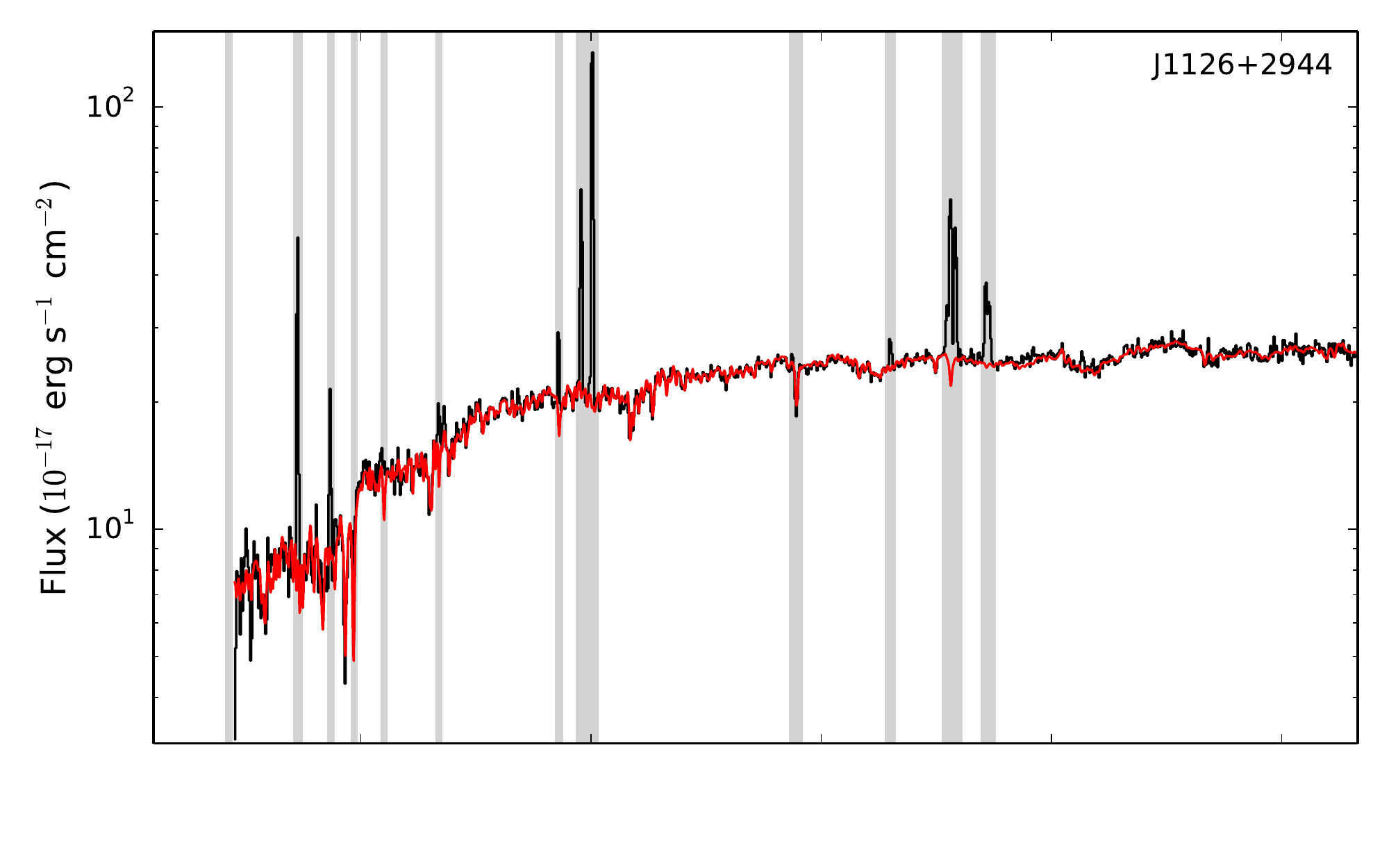} &
\includegraphics[width=3.5in]{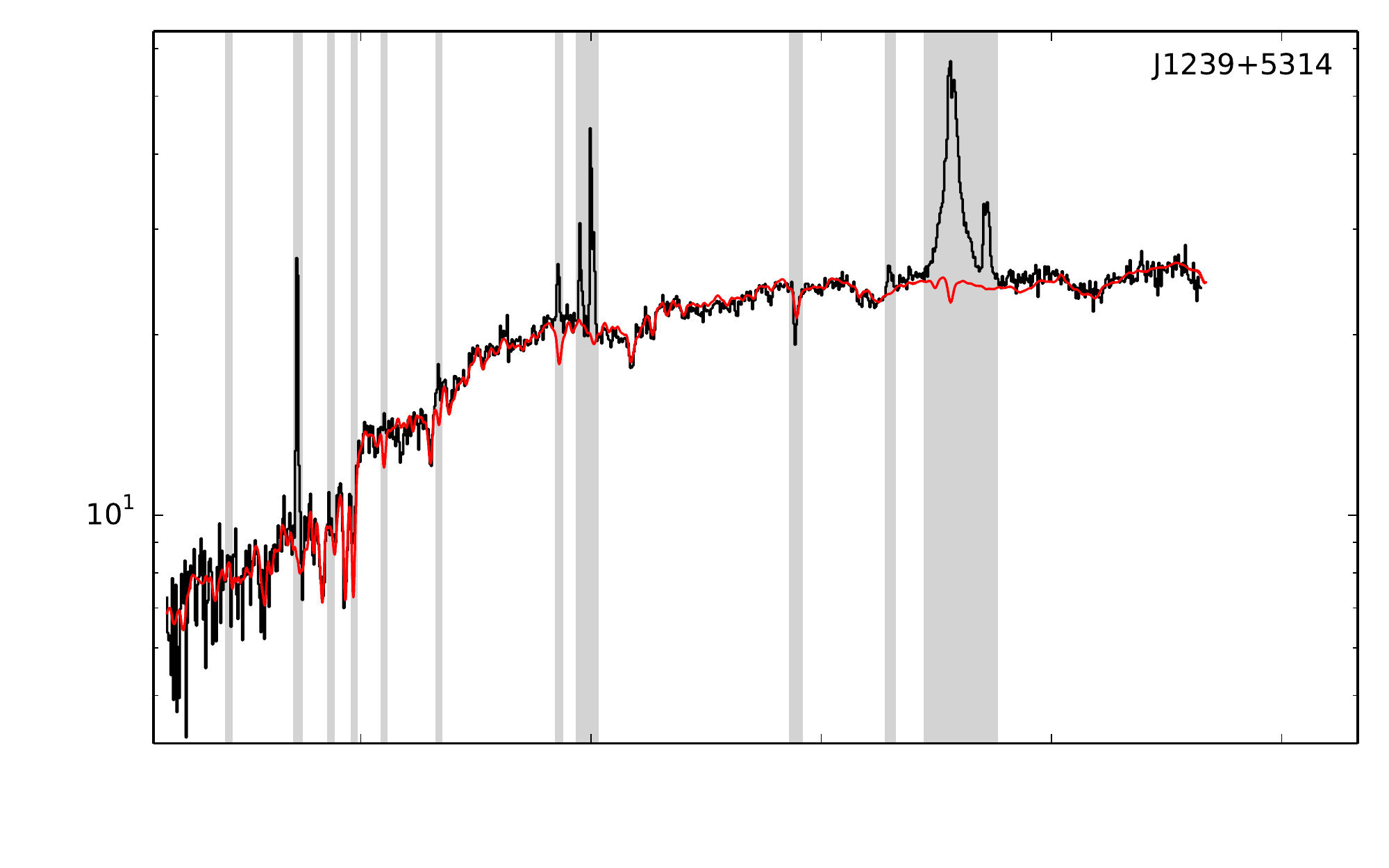} \\
\includegraphics[width=3.5in]{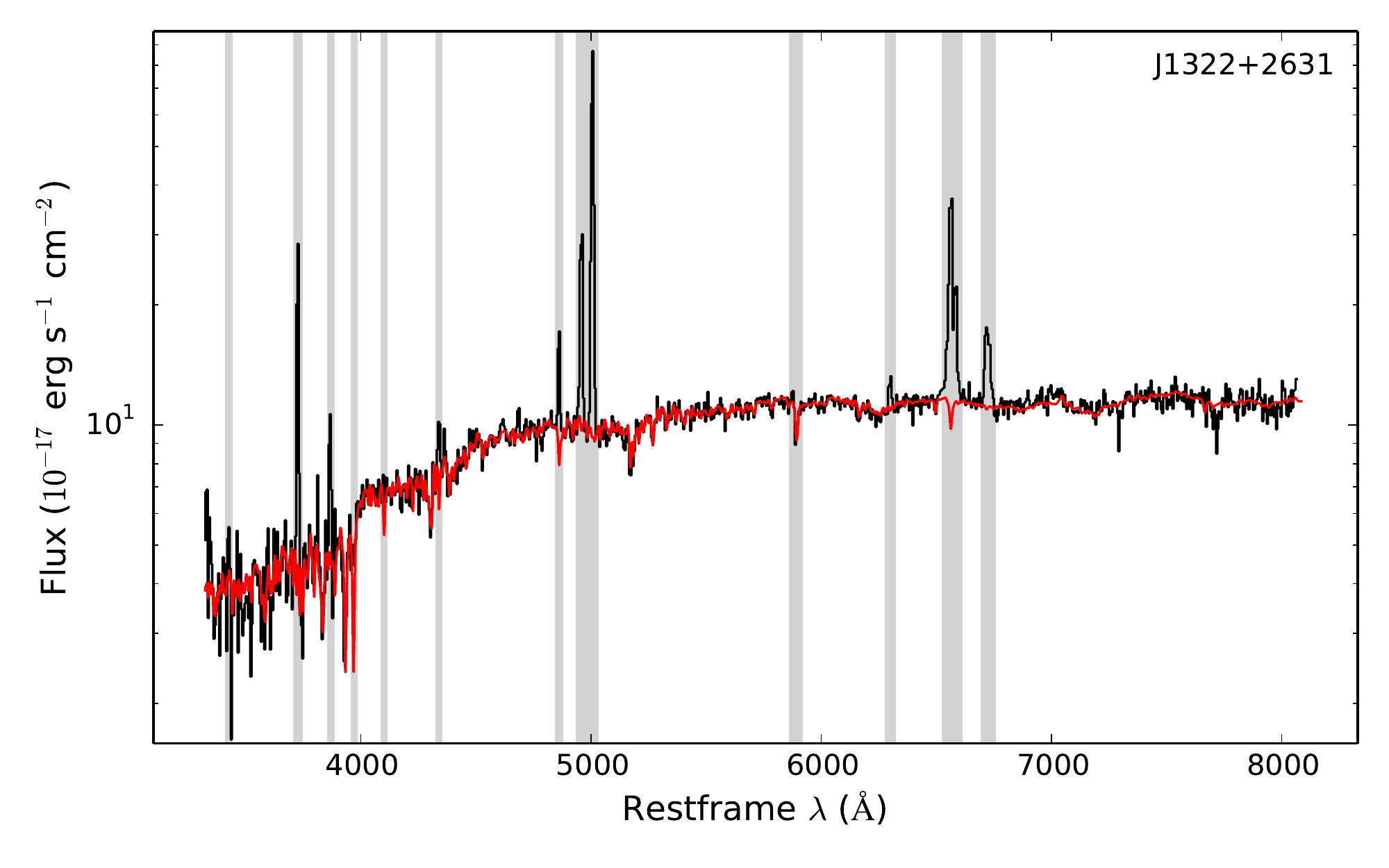} &
\includegraphics[width=3.5in]{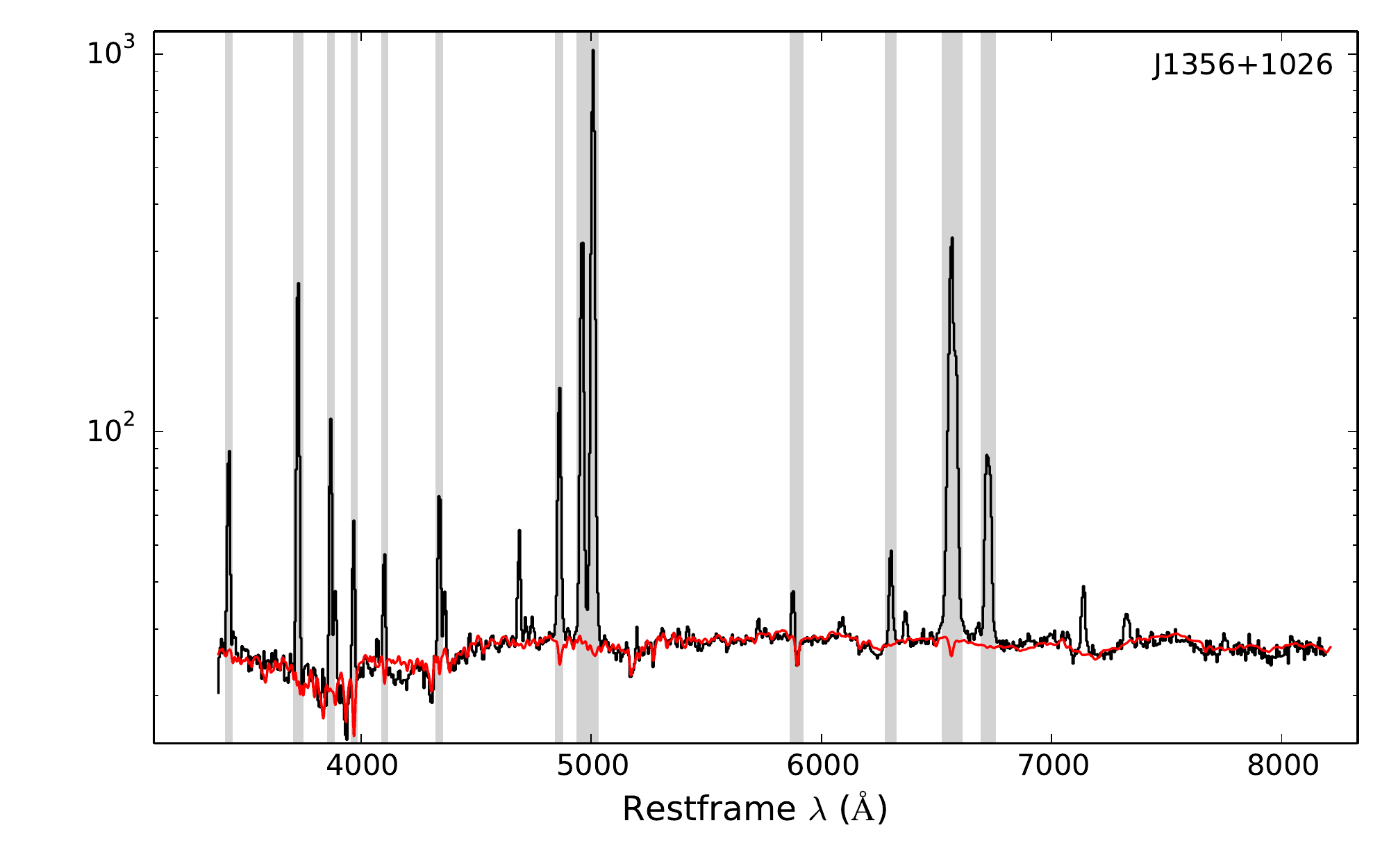}
\end{array} $
\caption{\footnotesize{Rest-frame SDSS fiber spectra of the \mergsz~merging systems (black, solid lines). The flux-weighted sum of the best-fit combination of 45 stellar bases is shown by the red, solid line. Spectral regions that were masked during the fitting (based on emission lines and bad pixels) are shown with gray shading. All spectra are shown over the same rest-frame wavelength range chosen so as to display the full SDSS spectrum for each source.}}
\label{fig:STARLIGHT}
\end{figure*}

Figure \ref{fig:plot_lrt_sdss} compares the values of \SFR~and \Mstar~from the UV-based SED method against those of the optical-based SDSS method. From the left panel of Figure \ref{fig:plot_lrt_sdss}, we see that the values of \SFRlrt~are well-correlated with those of \SFRsdss, consistent within their uncertainties and with an average offset from unity of \SFRoffset$\%$. For comparison, with a substantially larger sample \citet{Lee:2009} find that, for \SFR~above $0.1$~\SFRu~(which is the case for all measurements of \SFR~in our \mergsz~merger systems, regardless of method), the average offset from unity is $30\%$. The agreement within $6\%$ indicates the robustness of our measurements. From the right panel of Figure \ref{fig:plot_lrt_sdss}, we see that the values of \Mlrt~are well-correlated with those of \Msdss, consistent within their uncertainties and with an average offset from unity of \Moffset$\%$. Therefore, the differences between \sSFRlrt~and \sSFRsdss~are dominated by the differences in their respective \SFR~values.

The UV-based SED method is sensitive to longer \SF~timescales of $100$ Myr compared to the optical-based SDSS method sensitivity to timescales of $10$ Myr \citep{Lee:2009}. In our subsequent analysis, we adopt the \SFR~and \Mstar~estimates from SED modeling (\SFRlrt~and \Mlrt) since they do not involve applying an aperture correction, they account for the presence of an AGN continuum, the broad-band photometry covers a much larger range in energy, and we have measurements for all \mergsz~systems. \SFRlrt, \Mlrt, and \sSFRlrt~are shown in Table \ref{tab:SFR}. For completeness, we also include in Table \ref{tab:SFR} \SFRsdss, \Msdss, and \sSFRsdss.

\begin{deluxetable*}{ccccccccc}
\tabletypesize{\footnotesize}
\tablecolumns{9}
\tablecaption{Host galaxy properties derived from broadband and optical spectral modeling}
\tablehead{
\colhead{SDSS Name \vspace*{0.05in}} &
\colhead{\age} &
\colhead{\ebvagnlrt} &
\colhead{\SFRlrt} &
\colhead{\Mlrt} &
\colhead{\sSFRlrt} &
\colhead{\SFRsdss} &
\colhead{\Msdss} &
\colhead{\sSFRsdss} \\
\colhead{$-$ \vspace*{0.05in}} &
\colhead{(log[\ageu])} &
\colhead{(\ebvu)} &
\colhead{(\SFRu)} &
\colhead{(log[\Mu])} &
\colhead{(log[\sSFRu])} &
\colhead{(\SFRu)} &
\colhead{(log[\Mu])} &
\colhead{(log[\sSFRu])} \\
\colhead{(1)} &
\colhead{(2)} &
\colhead{(3)} &
\colhead{(4)} &
\colhead{(5)} &
\colhead{(6)} &
\colhead{(7)} &
\colhead{(8)} &
\colhead{(9)}
}
\startdata
J0841+0101 \vspace*{0.03in} & $9.89^{0.20}_{0.36}$ & $8.8_{5.1}^{9.8}$ & $2.9_{1.7}^{3.2}$ & $10.60_{0.42}^{0.34}$ & $-10.14_{0.43}^{0.35}$ & $6.1_{3.5}^{6.8}$ & $10.60_{0.38}^{0.33}$ & $-9.81_{0.37}^{0.33}$ \\
J0952+2552 \vspace*{0.03in} & $>6.00$\tablenotemark{a} & $0.1_{0.0}^{0.0}$ & $2.3_{0.2}^{0.2}$ & $12.21_{0.05}^{0.05}$ & $-11.86_{0.05}^{0.04}$ & - & - & - \\
J1126+2944 \vspace*{0.03in} & $10.03^{0.16}_{0.26}$ & $0.4_{0.3}^{1.1}$ & $0.6_{0.5}^{1.6}$ & $11.04_{0.74}^{0.57}$ & $-11.23_{0.50}^{0.52}$ & $0.6_{0.5}^{1.6}$ & $11.02_{0.61}^{0.55}$ & $-11.22_{0.58}^{0.54}$ \\
J1239+5314 \vspace*{0.03in} & $9.91^{0.20}_{0.39}$ & $1.4_{1.0}^{3.2}$ & $6.0_{4.2}^{13.6}$ & $11.81_{0.59}^{0.53}$ & $-11.03_{0.51}^{0.51}$ & $7.5_{5.3}^{17.0}$ & $11.84_{0.46}^{0.49}$ & $-10.97_{0.62}^{0.54}$ \\
J1322+2631 \vspace*{0.03in} & $9.98^{0.18}_{0.30}$ & $4.0_{3.2}^{12.5}$ & $1.4_{1.1}^{4.3}$ & $10.99_{0.79}^{0.63}$ & $-10.85_{0.74}^{0.63}$ & $0.4_{0.3}^{1.3}$ & $10.90_{0.78}^{0.63}$ & $-11.27_{0.77}^{0.63}$ \\
J1356+1026 \vspace*{0.0in} & $10.08^{0.17}_{0.28}$ & $10.2_{1.0}^{1.0}$ & $11.9_{1.2}^{1.2}$ & $11.20_{0.05}^{0.04}$ & $-10.12_{0.05}^{0.04}$ & - & - & - 
\enddata
\tablecomments{Column 1: abbreviated SDSS galaxy name; Column 2: mass-weighted mean age of the stellar populations from optical spectra modeling; Column 3: color excess of the AGN template from broadband SED modeling; Columns 4-6: star formation rate, stellar mass, and specific star formation rate from broadband SED modeling; Column 7-9: star formation rate, stellar mass, and specific star formation rate from optical spectral modeling.}
\tablenotetext{a}{Poorly constrained due to the presence of \feii~emission.}
\label{tab:SFR}
\end{deluxetable*}

\subsection{Stellar Ages}
\label{sec:spec}

The SDSS fiber spectra has observed spectral coverage of $3800-9200$~\AA, and the rest-frame spectral coverage of our sample ranges from $3440-8340$~\AA~(lowest redshift system) to $2840-6870$~\AA~(highest redshift system). We used \strlt~\citep{Fernandes:2004} to model the fiber spectra with synthesized stellar templates while masking resolved emission lines. Over the rest-frame wavelength coverage of the SDSS, the physical components that may contribute to the spectra consist of the host galaxy stellar continuum, AGN continuum, and broad/narrow emission lines. However, as shown in Section \ref{sec:sed} an AGN component is only significant at optical wavelengths in \typeI~AGN. Therefore, AGN continuum and broadened \feii~pseudo-continuum components are also included for the \typeI~AGN (\galaxytwo). Section \ref{sec:sed} also shows that reprocessed emission from dust only becomes significant at IR wavelengths while scattered AGN continuum light only becomes significant at UV wavelengths. Therefore, these components are not considered in our optical spectral modeling.

For the stellar populations, we used a library of 45 bases from \citet{Bruzual:Charlot:2003} that span three different metalicities (\metal~$=[0.004,0.02,0.05]$) and 15 different ages (\age~$=[0.001-13]\times10^{9}$yr). We applied a freely varying extinction curve \citep{Cardelli:1989} with \rv~$=3.1$ to the stellar bases. To include the AGN continuum for \galaxytwo, we added to the library a power-law base defined by $F_{\lambda}=10^{20}(\lambda/4020$\AA$)^{-\alpha}$ erg s$^{-1}$ cm$^{-2}$ $\rm{\AA}^{-1}$ as in \citet{Mezcua:2011}. For the power-law index $\alpha$, we investigated the results using six discrete values: $\alpha=-[0.5,-1.0,-1.5,-2.0,-2.5,-3.0]$. We applied a separate freely varying extinction curve to the power-law base representing the AGN. When varying $\alpha$ over the six values, we find that the values of \age~are all consistent within their respective uncertainties. Therefore, we adopted the value of $\alpha=-1.5$ as in Section \ref{sec:additional}. We also added to its library a base of empirically derived \feii~pseudo-continuum created from the templates of \citet{Veron_Cetty:2004} and \citet{Tsuzuki:2006}. When including the \feii~component, the best-fit combination of stellar bases is dominated by the youngest population (\age~$=10^{6}$ yrs). Therefore, we consider this \age~value to be a lower limit. The optical fiber spectra and best fit models are shown in Figure \ref{fig:STARLIGHT}. We then computed mean stellar ages (\age) as the mass-weighted average age of the individual stellar populations. Values of \age~are listed in Table \ref{tab:SFR}.

\subsection{Building the Control Sample}
\label{sec:build_control}

To understand the values of \SFR, \Mstar, and \age~for our sample within the context of the AGN population, we build control samples for each merger system. To do so, we matched these \mergsz~merger systems to AGN (selected based on the same narrow emission line ratios from Section \ref{sec:sample}) in the \mpa~galaxy catalogue on values of \Mstar, \zsdss, and \loiii. We included galaxies in control samples if matches on \Mstar, \zsdss, and \loiii~are within the thresholds of $20\%$, $20\%$, and $50\%$, respectively, as used in \citet{Nevin:2017}. This ensures that control samples for the four AGN originally selected from the \mpa~catalog (\galaxyone, \galaxythree, \galaxyfour, and \galaxyfive) include at least 10 sources. The other two (\galaxytwo~and \galaxysix) are both classified as quasars (\typeI~and \typeII, respectively) and therefore are more luminous than the other four. \galaxysix~requires matchhes to \loiii~to be within $75\%$ to provide at least $10$ sources in the control sample. \galaxytwo~requires matches to \loiii~to be within $100\%$, plus an increase to $50\%$ for matches on \Mstar~and \zsdss~to provide at least $10$ sources in the control sample. The final numbers in each control sample are \ctrlszone~(\galaxyone), \ctrlsztwo~(\galaxytwo), \ctrlszthree~(\galaxythree), \ctrlszfour~(\galaxyfour), \ctrlszfive~(\galaxyfive), and \ctrlszsix~(\galaxysix). We ran the \lrt~SED modeling procedure (using the same photometric surveys as in Section \ref{sec:sed}) and the \strlt~spectral modeling procedure (using the SDSS optical fiber spectra as in Section \ref{sec:spec}) on all galaxies from each control sample. Finally, we computed the average values for each control sample to obtain \SFRctrl, \Mctrl, and \agectrl.

As discussed in Section \ref{sec:additional}, light from scattering, extended NLRs, and AGN-heated dust can contribute to the photometry beyond the SED modeling capabilities. While we have shown that the effects of emission lines and heated dust are likely to be small, scattered AGN continuum emission can elevate our measurements of \sSFRlrt~to artificially high levels. This effect will also occur among the AGN in the control samples and will affect comparisons if the average AGN luminosity of the control sample (which dominates the scattering contribution) is significantly different from that of the AGN to which it is matched. Based on \loiii, we find that the AGN luminosities of the \mergsz~merger systems are greater than the average of the matched control samples by factors of \galfourAGNctrlfac$-$\galtwoAGNctrlfac. This implies that the over-estimates of \sSFRlrt~in the \mergsz~merger systems are more significant than in the control samples in all cases. The implications of this effect are considered in Sections \ref{sec:correlations} and \ref{sec:control}.

\section{Image Analysis}
\label{sec:image_analysis}

In this Section we describe our analysis of the imaging data for the \mergsz~merging systems. In Section \ref{sec:galfit_model} we fit two-dimensional parametric models to the \Hb-band images to model the spatial distribution of the stars and assess the presence of morphological disturbances. In Section \ref{sec:color_gradients} we characterize the radial color gradients to quantify the spatial distribution of young stellar populations.

\subsection{Morphological Disturbances}
\label{sec:galfit_model}

From Section \ref{sec:sed} we see that all \mergsz~merger systems have small flux contributions in the \Hb-band (peak wavelength of $15450$~\AA~and $FWHM=2900$~\AA) from scattered light ($<$\galonescatHfrac$\%$), and no measurable contribution from emission lines and AGN-heated dust. We see that five of the systems (\galaxyone, \galaxythree, \galaxyfour, \galaxyfive, and \galaxysix) have small flux contributions in the \Hb-band from AGN light (\galsixAGNHfrac$-$\galfiveAGNHfrac$\%$). While the remaining system (\galaxytwo) is a \typeI~AGN with a stronger AGN contribution (\galtwoAGNHfrac$\%$), the AGN flux is from the continuum and therefore is likely confined to the central four pixels (based on the \Hb-band PSF) so that it contributes negligibly to the global \Hb-band morphology. Therefore, the \Hb-band galaxy images are dominated by light from stellar continuum emission. This flux corresponds to stellar populations with spectral curves peaking at NIR wavelengths and hence relatively older stars \citep{Mannucci:2001} that represent the classical bulge components of galaxies.

\begin{figure*} $
\begin{array}{cc}
\vspace*{-0.0in} \hspace*{0.in} \includegraphics[width=0.5\textwidth]{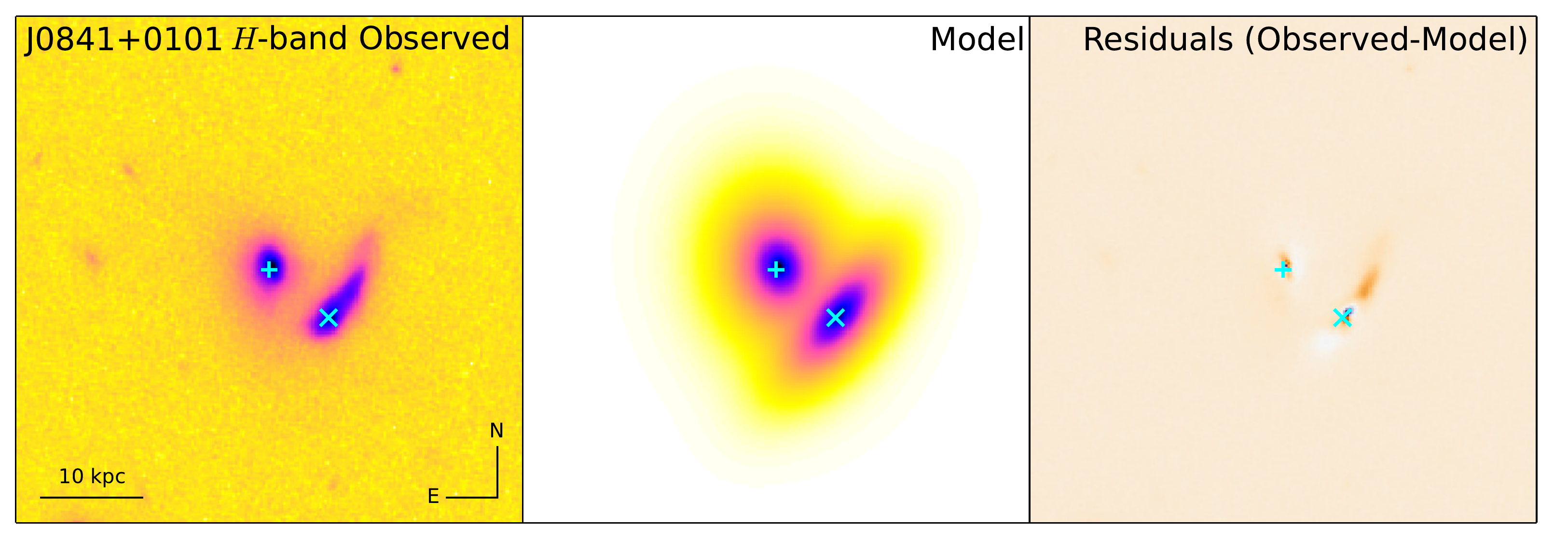} &
\vspace*{-0.0in} \hspace*{0.in} \includegraphics[width=0.5\textwidth]{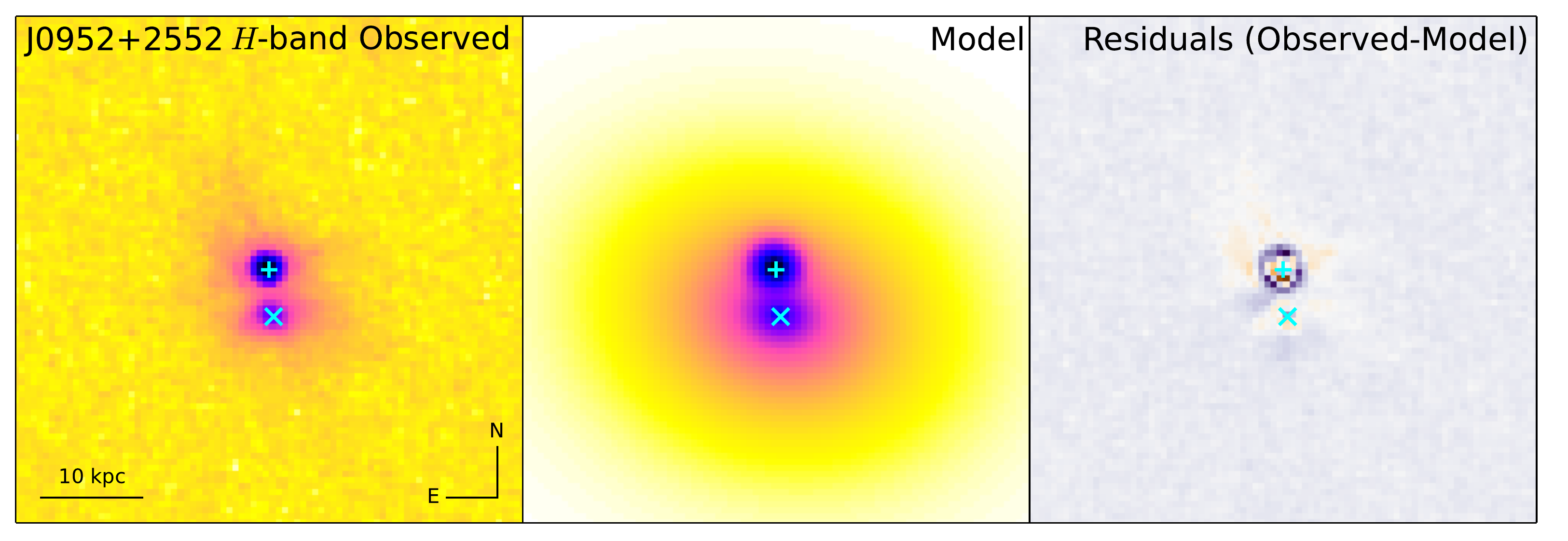} \\
\vspace*{-0.0in} \hspace*{0.in} \includegraphics[width=0.5\textwidth]{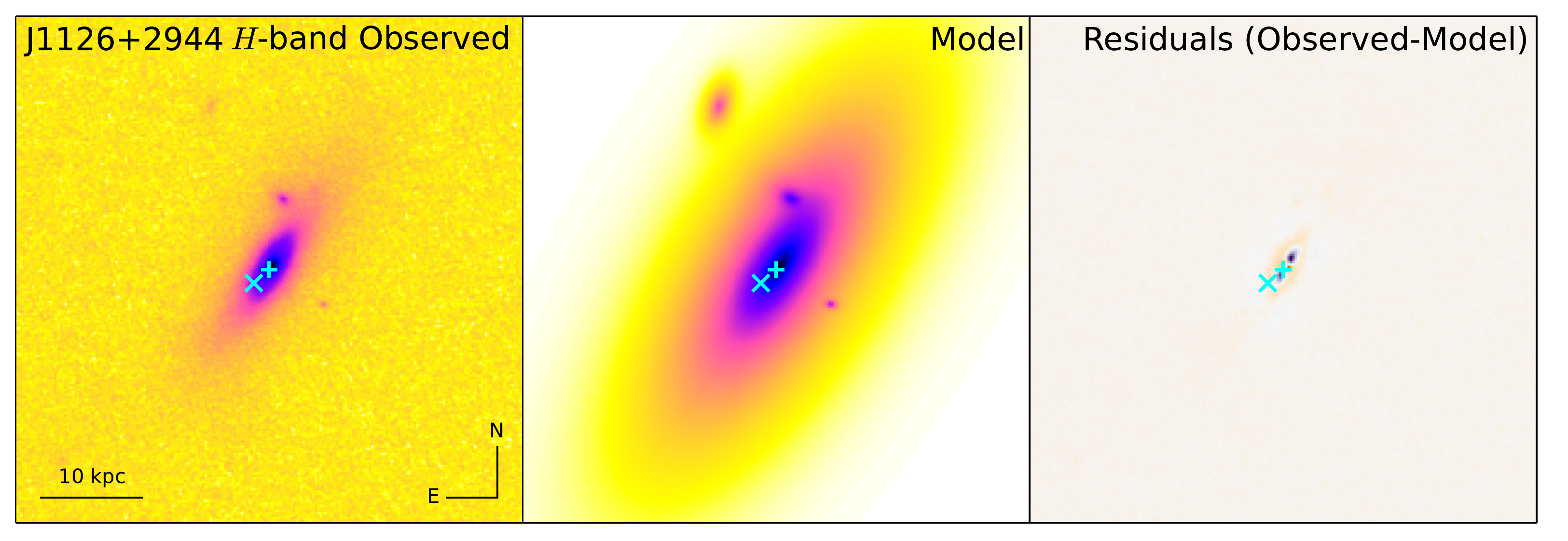} &
\vspace*{-0.0in} \hspace*{0.in} \includegraphics[width=0.5\textwidth]{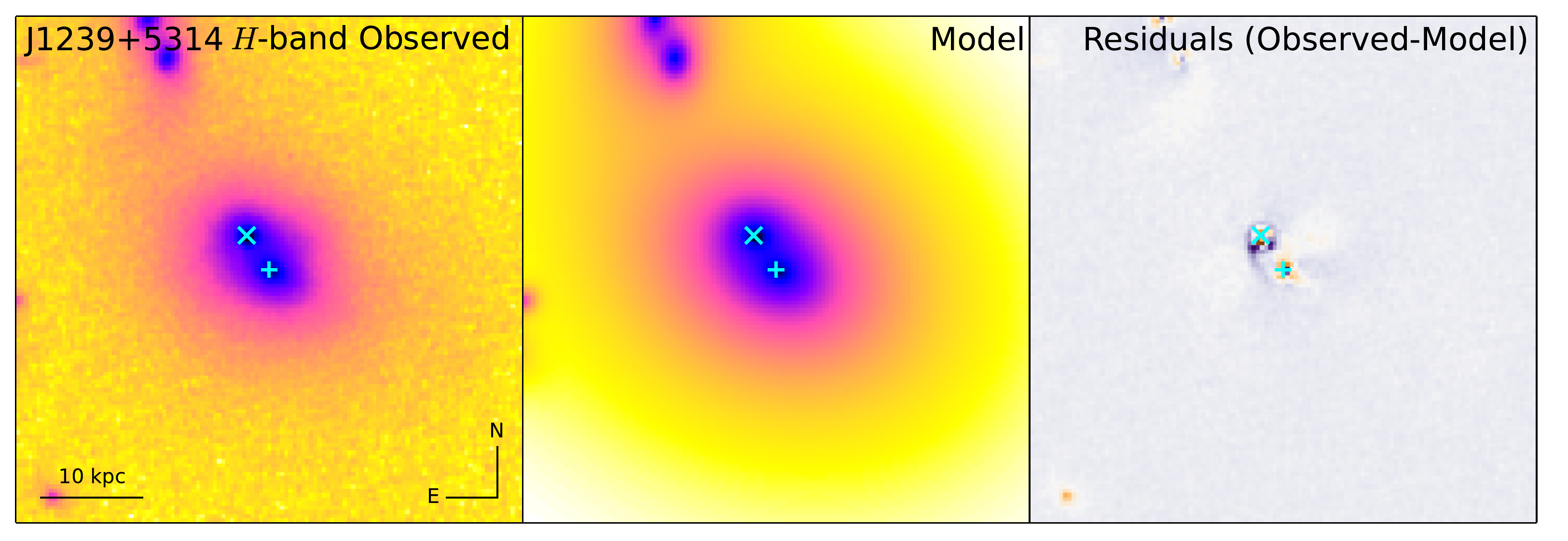} \\
\vspace*{-0.0in} \hspace*{0.in} \includegraphics[width=0.5\textwidth]{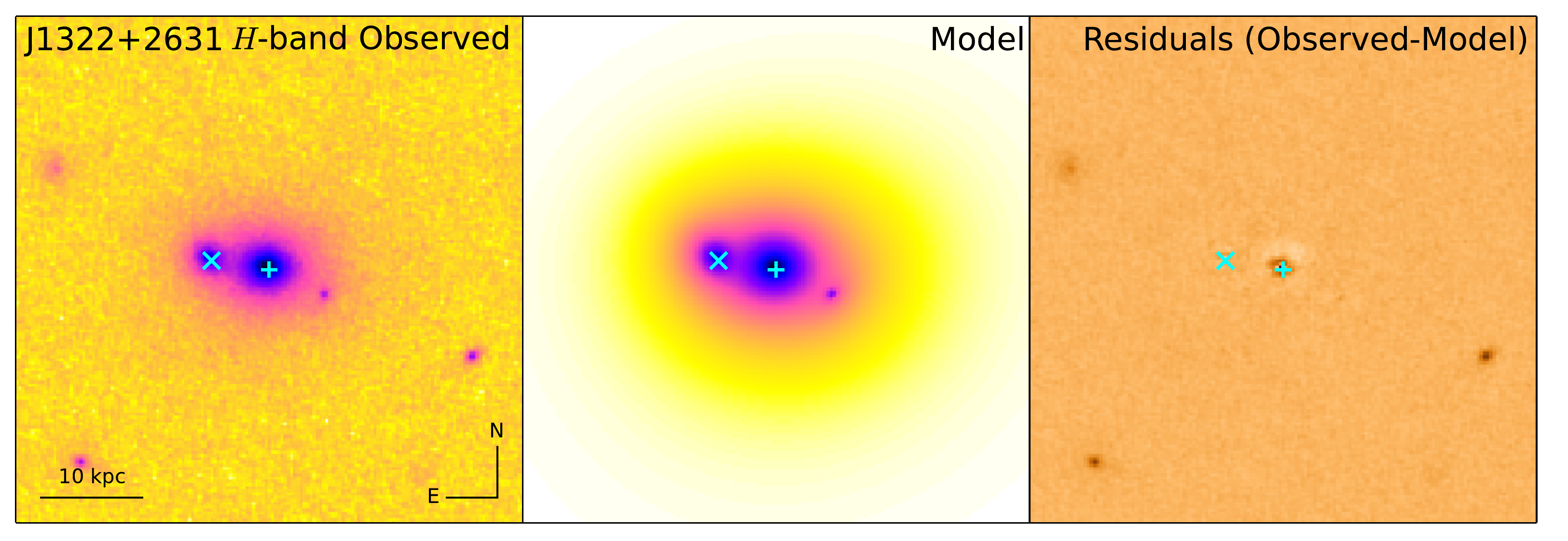} &
\hspace*{0.in} \includegraphics[width=0.5\textwidth]{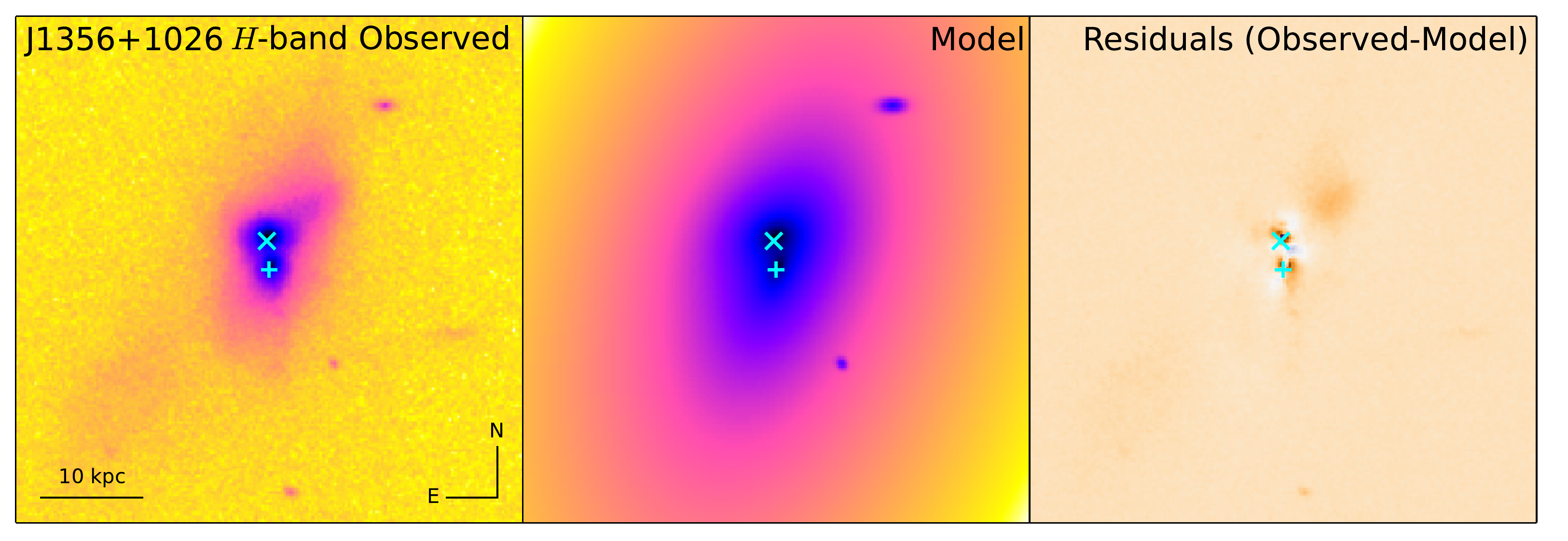}
\end{array}$
\caption{\footnotesize{From left to right: \Hb-band image, best \gf~model of the \Hb-band image, and the residuals obtained by subtracting the model from the \Hb-band image. All image panels are displayed over a \fitdim$\times$\fitdim~kpc FOV (\gf~fitting box). The cyan `+' and `x' represent the locations of the \prim~and \second~components of each merger, respectively. The component parameters from these models are analyzed in Section \ref{sec:results}.
}}
\label{fig:GALFIT}
\end{figure*}

To model the \Hb-band images of the merger systems, we fit each with a combination of multiple Sersic functions. The Sersic function \citep{Sersic:1968} is a reliable model for a wide range of stellar light radial profiles \citep{Graham:2005}, with the freely varying exponential parameter \n~(Sersic index) describing how light is concentrated around the peak. Smaller values of \n~tend toward more concentrated profiles with smooth peaks, while larger values of \n~tend toward more extended profiles with cuspy peaks. The Sersic function is widely used to model stellar bulges of galaxies, either in elliptical galaxies or the central regions of disk galaxies often with the special case of $n=4$ \citep{deVaucouleurs:1974}, though the case of \n~$=1$ has often been used to model the exponential profiles of edge-on disk galaxies \citep{Patterson:1940}. Since the utility of Sersic profiles has been evaluated on the classical morphologies of early- and late-type galaxies, systems that significantly deviate from those morphologies will not be adequately fit by this function. 

In \paperI, we demonstrated that the Sersic centroid fit, obtained with the galaxy image modeling program \gf~version 3.0.5 \citep{Peng:2010}, is a robust tracer of the peak \Hb-band brightness, regardless of the residuals at large radii from the centroid. However, in this work we are also interested in quantifying the global morphologies of the merger systems and the deviations from symmetry that may have arisen from the mergers. Therefore, we redo the analysis here to also include nearby sources so that we can accurately measure the residuals at all radii. First, we run \se~\citep{Bertin:Arnouts:1996} on the \Hb-band images to generate a base list of all the significantly detected ($>3\sigma$) sources. From this list, we fit Sersic components to all of the detected sources within a \fitdim$\times$\fitdim~kpc field-of-view (FOV) centered on the SDSS J2000 right ascension and declination of the galaxy, plus a fixed, uniform sky background estimated from a source-free region. The choice of a \fitdim$\times$\fitdim~kpc FOV allows all nearby contaminating sources to be included in the \Hb-band fitting box and modeled for all \mergsz~galaxies. We eliminate some Sersic components from the final fit if they do not centroid on the correct \se~position. We test fitting PSFs to the systems, finding that it is necessary only for the faint Southeast source in \galaxythree~(see \paperI~for details). While \galaxytwo~hosts a \typeI~AGN based on its SDSS spectrum, our analysis in \paperI~determined that adding a PSF component does not improve the quality of the fit at a statistically significant level. The \Hb-band images and results of the \gf~modeling are shown in Figure \ref{fig:GALFIT}. As a form of quantifying the level of morphological disturbances in each system we also measured the rotational asymmetry, \Arot, of each galaxy following the procedure defined in \citet{Conselice:2009}.

\begin{figure*} $
\begin{array}{c}
\vspace*{-0.08in} \hspace*{0.5in} \includegraphics[width=0.8\textwidth]{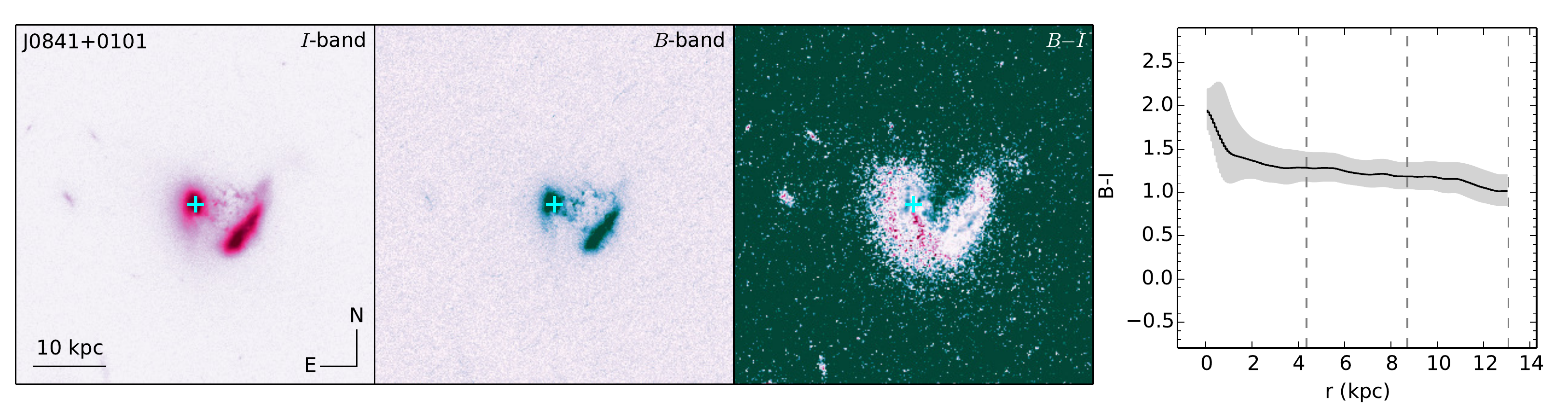} \\
\vspace*{-0.08in} \hspace*{0.5in} \includegraphics[width=0.8\textwidth]{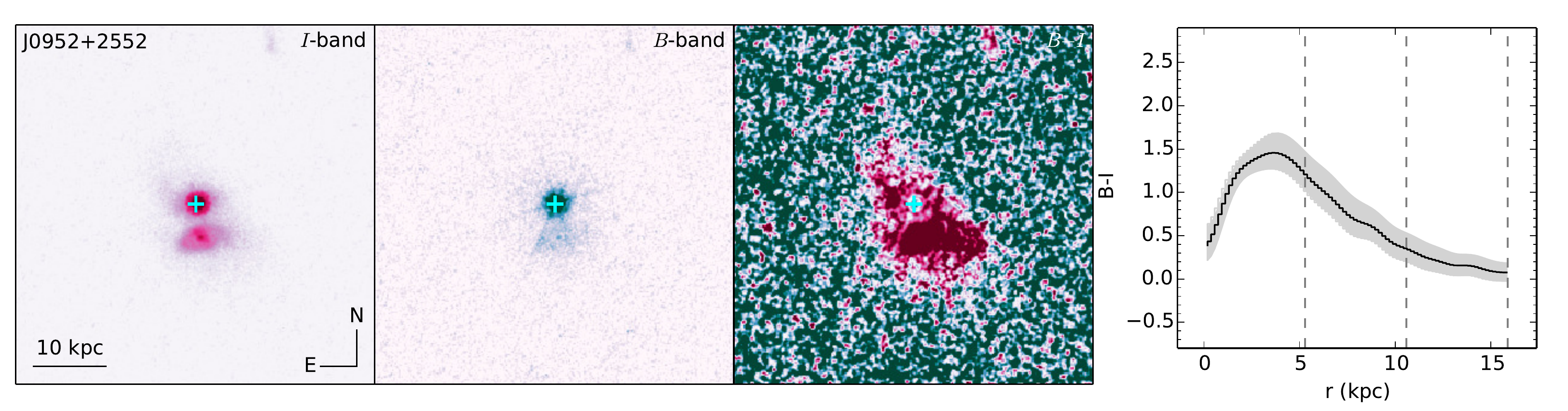} \\
\vspace*{-0.08in} \hspace*{0.5in} \includegraphics[width=0.8\textwidth]{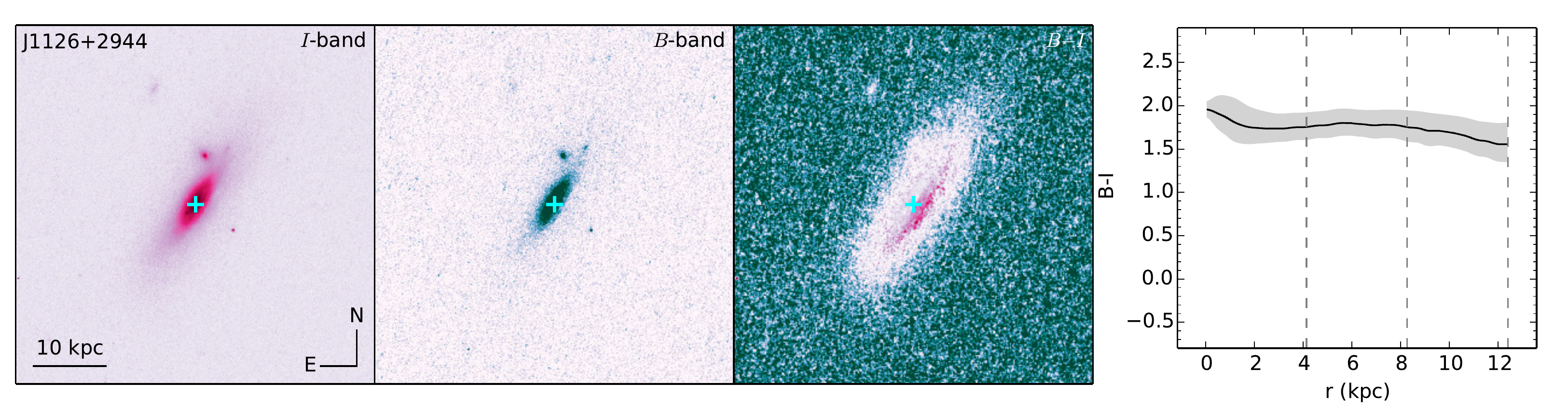} \\
\vspace*{-0.08in} \hspace*{0.5in} \includegraphics[width=0.8\textwidth]{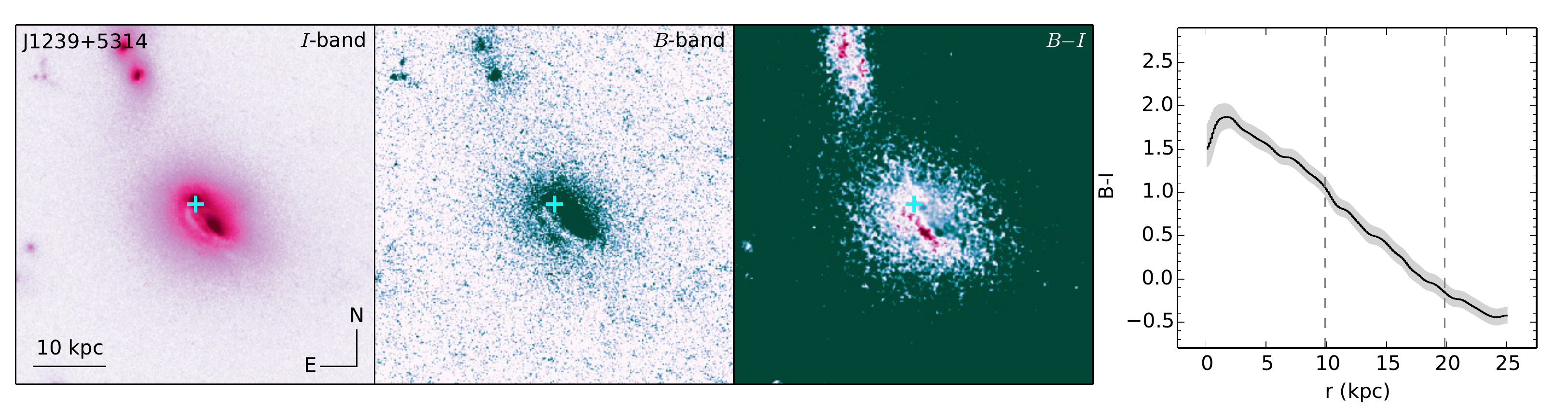} \\
\vspace*{-0.08in} \hspace*{0.5in} \includegraphics[width=0.8\textwidth]{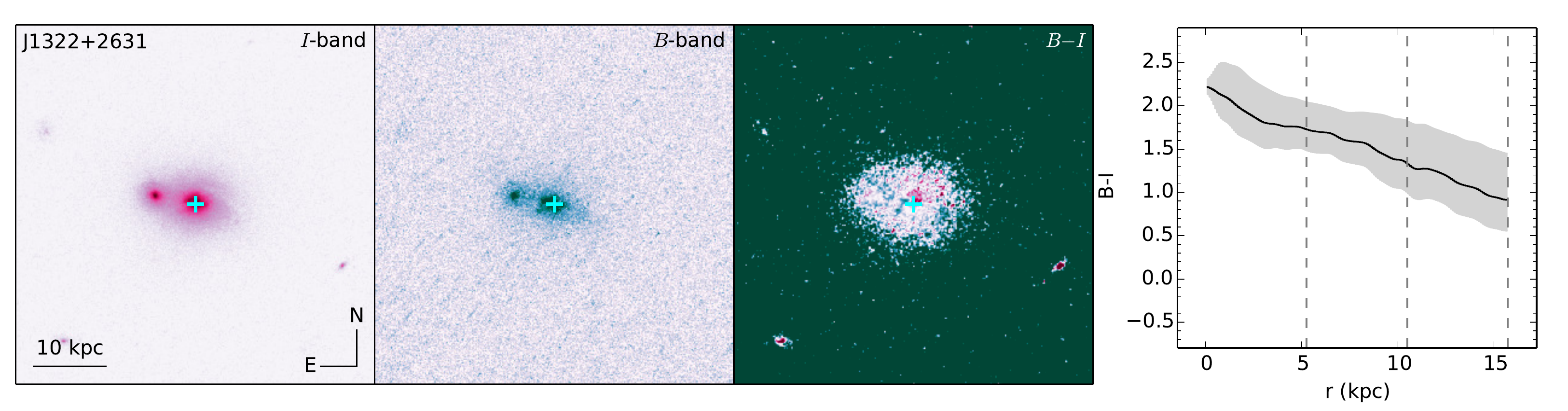} \\
\hspace*{0.5in} \includegraphics[width=0.8\textwidth]{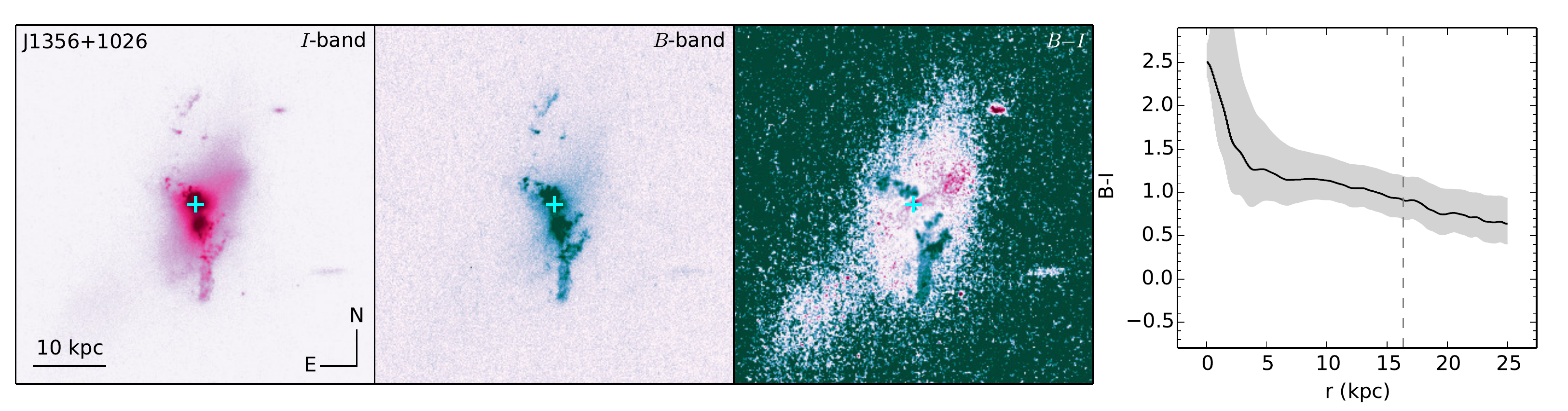}
\end{array}$
\caption{\footnotesize{From left to right, the three image panels show the following: \Ib-band image, \Bb-band image and the \BI~color image with their color scales combined. All image panels are displayed over a \fitdim$\times$\fitdim~kpc FOV. The cyan cross represents the location of the primary components of each merger. The far right panel shows the \Bb-\Ib~color as a function of physical distance from the \prim~Sersic component out to \ellRefacb$\times$\recomb, with the upper and lower \BIerrfac$\sigma$ uncertainty bounds indicated by the gray, shaded regions. Starting from $0$ kpc, the vertical, black dashed lines represent the $1$, $2$, and $3$\recomb~radii (for radii of $<25$ kpc). Note that, while the radial profiles at small radii have varying slopes due to the presence of AGN continuum emission in some cases, the slopes are negative at large radial distance in all systems. The average \BI~colors and radial slopes are analyzed in Section \ref{sec:results}.}}
\label{fig:HST_color}
\end{figure*}

The two Sersic components nearest the SDSS coordinates of the galaxy (in the \Hb-band reference frame) are considered to be the nuclear components of the merger, with the brightest (based on the \Hb-band Sersic component magnitudes) referred to as the \prim~(component 1) and the other as the \second~(component 2). In Table \ref{tab:sersic} we list the half-light radii (\reone~and \retwo) and Sersic indices (\none~and \ntwo) for the \prim~and \second~components, along with \Arot. In Section \ref{sec:results} we use these parameters to test if the stellar morphologies are correlated with \SF~properties.

\subsection{Color Gradients}
\label{sec:color_gradients}

From Section \ref{sec:sed} we see that all \mergsz~merger systems have small flux contributions in the \Ib-band (peak wavelength of $8353$~\AA~and $FWHM=2555$~\AA) and \Bb-band (peak wavelength of $4320$~\AA~and $FWHM=695$~\AA) from scattered light (\galthreescatIfrac$-$\galonescatBfrac$\%$), and emission lines ($<$\galaxyoneHSTIemflx$\%$) and no measurable contribution from AGN-heated dust. As with the \Hb-band, five of the systems (\galaxyone, \galaxythree, \galaxyfour, \galaxyfive, and \galaxysix) have small flux contributions in the \Ib- and \Bb-bands from AGN light ($<$\galthreeAGNBfrac$\%$). Also as with the \Hb-band, while the remaining merger system (\galaxytwo) is a \typeI~AGN with a stronger AGN contribution (\galtwoAGNIfrac$-$\galtwoAGNBfrac$\%$), the AGN flux spatial distribution is described by the \Ib- and \Bb-band PSFs and so contributes negligibly to the global \Ib- and \Bb-band morphologies. Therefore, the \Ib- and \Bb-band galaxy images are dominated by light from stellar continuum emission. This flux corresponds to stellar populations with spectral curves peaking at red optical wavelengths (\Ib-band) and blue optical/NUV wavelengths (\Bb-band). These signatures represent relatively old and young stellar populations for the \Ib- and \Bb-bands, respectively \citep{Kaviraj:2007} Thus, the difference between the \Ib- and \Bb-band galaxy images (\BI) can reveal spatial distributions of star-forming regions.

The \Ib- and \Bb-band images are the optimal combination for color maps as they have the same pixel scales and the best spatial resolution (seeing FWHMs of $0\farcs074$ and $0\farcs070$ for the \Ib- and \Bb-bands, respectively). To construct \BI~maps, we first register both the \Ib- and \Bb-band images (input images) to the same coordinate system as the \Hb-band image (reference image). We use the astrometric procedure from \citet{Barrows:2016}, optimized for a small FOV, to register the two images. To combine the images, we follow a procedure similar to the one used in \citet{Shangguan:2016} that consists of convolving each image with a Gaussian kernel so that the two images are at a common resolution of FWHM$=0\farcs080$ and replacing any negative pixel values with interpolated values based on the Gaussian kernel. We then divide pixel values in the \Ib-band image by those in \Bb-band image, and compute \BI~using the photometric zeropoints. Figure \ref{fig:HST_color} shows the images of the six mergers in the \Ib- and \Bb-band images, along with the \BI~color images.

Finally, we use the \el~task in \iraf~to measure the counts and semi-major axes of isophotes from the \BI~color images in linear steps of one pixel from the location of the \prim~stellar core out to \ellRefacb$\times$\recomb, where \recomb$=$\reone+\retwo~(but no larger than the \gf~fitting box). We plot \BI~against isophotal semi-major axis in the far right panel of Figure \ref{fig:HST_color}. The upper and lower \BIerrfac$\sigma$ uncertainty bounds are calculated from the isophotal intensity uncertainties generated by \el. The average \BI~value within \recomb, \BIavg, and radial slope of \BI~outside of \recomb~calculated from a linear regression of the data, \BIgrad, are listed in Table \ref{tab:sersic}. In Section \ref{sec:results} we use these parameters to determine how the distributions of \SF~compare with the overall populations of early-type, late-type and peculiar or merging galaxies.

\section{Results}
\label{sec:results}

In this section we describe the main results from the analyses presented in Sections \ref{sec:spec_analysis} and \ref{sec:image_analysis}. Our aim is to reveal connections between \SF~and the merger processes. Specifically, we describe how the \SF~is spatially distributed within the merging systems (Section \ref{sec:radii}), describe how properties of the stellar populations correlate with global properties of the galaxies (Section \ref{sec:correlations}), and describe how the \mergsz~merger systems compare to their control samples (Section \ref{sec:control}).

\subsection{Star Formation Is on a Global Scale}
\label{sec:radii}

While early-type galaxies display relatively shallow radial changes in color \citep{Hunt:1997,Bartholomew:2001,Tamura:Ohta:2003,Cantiello:2005}, peculiar (interacting or merging) galaxies are known to produce a wide range of color gradients likely caused by spatial differences in \SF~and the presence of dust \citep{Taylor:2005}. The radial color gradients outside of \recomb~for the \mergsz~mergers, \BIgrad, are shown in Table \ref{tab:sersic}. 

\begin{deluxetable*}{cccccccccc}
\tabletypesize{\footnotesize}
\tablecolumns{10}
\tablecaption{Host galaxy properties derived from NIR and optical image modeling.}
\tablehead{
\colhead{SDSS Name \vspace*{0.05in}} &
\colhead{\reone} &
\colhead{\retwo} &
\colhead{\none} &
\colhead{\ntwo} &
\colhead{\Arot} &
\colhead{\BIavg} &
\colhead{\BIgrad} &
\colhead{\BVavg} &
\colhead{\BVgrad} \\
\colhead{$-$ \vspace*{0.05in}}  &
\colhead{(\physsepu)} &
\colhead{(\physsepu)} &
\colhead{$-$} &
\colhead{$-$} &
\colhead{$-$} &
\colhead{(\magu)} &
\colhead{(\gradu)} &
\colhead{(\magu)} &
\colhead{(\gradu)} \\
\colhead{(1)} &
\colhead{(2)} &
\colhead{(3)} &
\colhead{(4)} &
\colhead{(5)} &
\colhead{(6)} &
\colhead{(7)} &
\colhead{(8)} &
\colhead{(9)} &
\colhead{(10)}
}
\startdata
J0841+0101 \vspace*{0.03in} & $1.94\pm0.01$ & $2.41\pm0.01$ & $2.29\pm0.01$ & $1.48\pm0.01$ & $ 0.98 $  & $ 1.44_{0.02}^{0.01} $ & $-0.026_{0.005}^{0.006}$ & $ 0.35_{0.01}^{0.01} $ & $-0.026_{0.005}^{0.005}$ \\
J0952+2552 \vspace*{0.03in} & $0.88\pm0.00$ & $4.42\pm0.12$ & $1.18\pm0.01$ & $2.16\pm0.05$ & $ 0.52 $  & $ 1.14_{0.02}^{0.02} $ & $-0.439_{0.032}^{0.027}$ & $ 0.01_{0.02}^{0.02} $ & $-0.439_{0.031}^{0.029}$ \\
J1126+2944 \vspace*{0.03in} & $3.79\pm0.02$ & $0.35\pm0.04$ & $2.40\pm0.01$ & - & $ 0.27 $  & $ 1.79_{0.02}^{0.01} $ & $-0.023_{0.006}^{0.006}$ & $ 0.21_{0.01}^{0.01} $ & $-0.023_{0.005}^{0.005}$ \\
J1239+5314 \vspace*{0.03in} & $3.10\pm0.04$ & $6.82\pm0.08$ & $3.46\pm0.02$ & $2.56\pm0.02$ & $ 0.74 $  & $ 1.53_{0.01}^{0.01} $ & $-0.106_{0.002}^{0.002}$ & $ -0.00_{0.01}^{0.01} $ & $-0.106_{0.002}^{0.002}$ \\
J1322+2631 \vspace*{0.03in} & $3.25\pm0.03$ & $1.99\pm0.03$ & $2.10\pm0.01$ & $1.79\pm0.02$ & $ 0.35 $  & $ 1.93_{0.03}^{0.03} $ & $-0.096_{0.011}^{0.011}$ & $ 0.54_{0.03}^{0.04} $ & $-0.096_{0.011}^{0.015}$ \\
J1356+1026 \vspace*{0.0in} & $1.99\pm0.02$ & $14.34\pm0.37$ & $2.01\pm0.01$ & $3.86\pm0.04$ & $ 0.76 $  & $ 1.28_{0.03}^{0.02} $ & $-0.009_{0.002}^{0.003}$ & $ 0.02_{0.03}^{0.02} $ & $-0.009_{0.002}^{0.003}$  
\enddata
\tablecomments{Column 1: abbreviated SDSS galaxy name; Columns 2-3: effective radii of the \prim~(\reone)~and \second~(\retwo) Sersic components; Column 4-5: indices of the \prim~(\none) and \second~(\ntwo) Sersic components; Column 6: rotational asymmetry index; Column 7: average of \BI~within \ellRefaca$\times$\recomb~of the \prim~Sersic component; Column 8: gradient of \BI~within \ellRefaca$-$\ellRefacb$\times$\recomb~of the \prim~Sersic component; Column 9: average of \BV~within \ellRefaca$\times$\recomb~of the \prim~Sersic component; Column 10: gradient of \BV~within \ellRefaca$-$\ellRefacb$\times$\recomb~of the \prim~Sersic component. \retwo~(Column 3) and \ntwo~(Column 5) are not measured for \galaxythree~(Section \ref{sec:galfit_model}).}
\label{tab:sersic}
\end{deluxetable*}

Based on their uncertainties, the values of \BIgrad~for all \mergsz~systems are consistent with being negative at $>3\sigma$ significance and hence the galaxies become bluer with increasing radial distance from the nucleus. The analysis in Section \ref{sec:color_gradients} shows that the \BI~colors in these systems are likely dominated by stellar light, and therefore the negative gradients are produced by younger stellar populations and enhanced \SF~at large radii compared to the nuclear regions. With the exception of \galaxytwo~(the system with the largest \Ib- and \Bb-band AGN contributions) this result is true even when including the inner \recomb~radii and is qualitatively similar to the galaxy colors presented for the late-stage merger sample of \citet{Shangguan:2016}.

\begin{figure}
\hspace*{-0.1in} \includegraphics[width=3.5in]{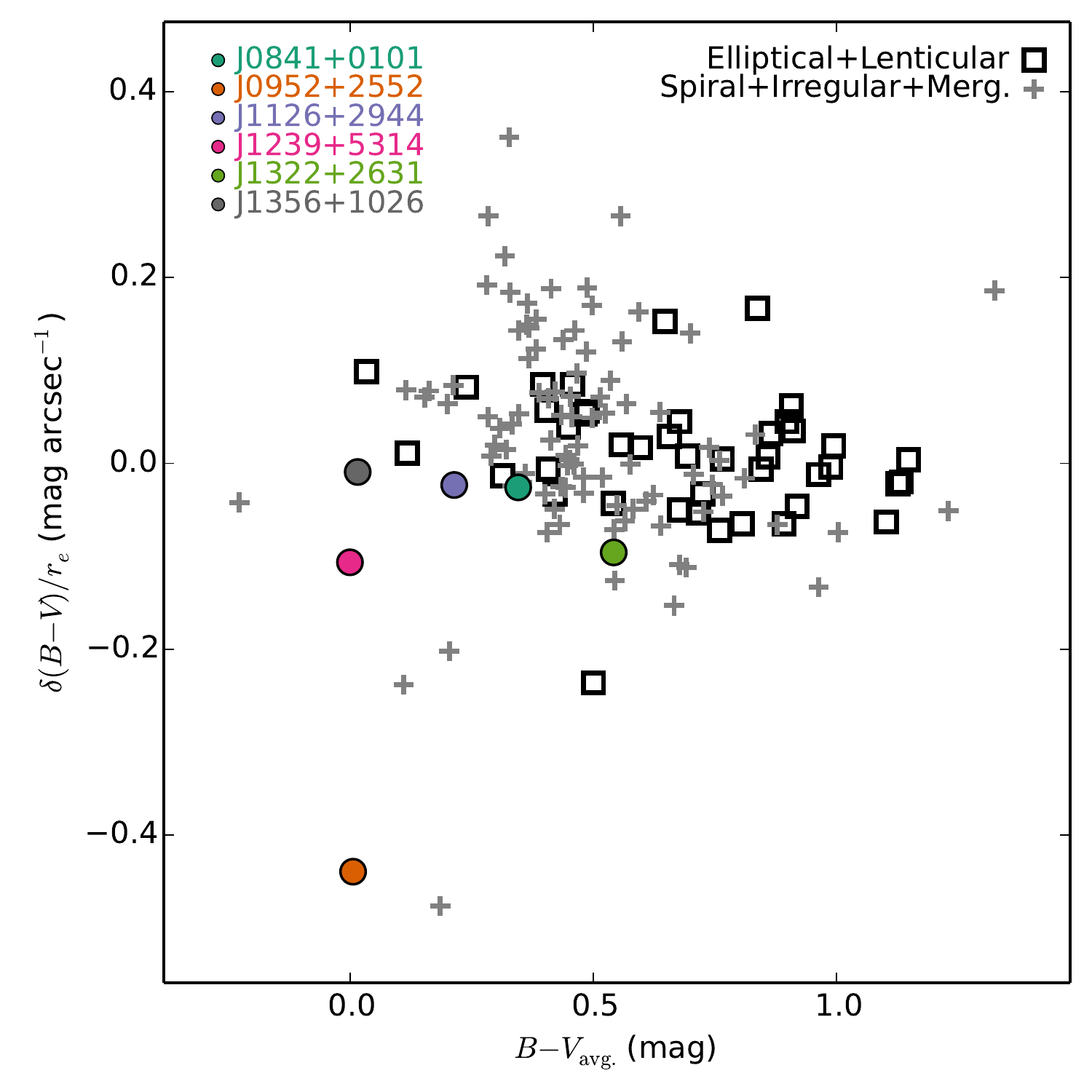}
\caption{\footnotesize{Radial \BV~color gradient (outside of \recomb) as a function of average $B-V$ (within \recomb). The \mergsz~systems from this work are shown as filled circles. Also shown are the elliptical and lenticular systems (black, open squares) and spiral, irregular and merging systems (gray crosses) from \citet{Taylor:2005}.}}
\label{fig:plot_age_color_avg_slope}
\end{figure}

To place our \mergsz~merger systems within the context of typical galaxy color gradients, in Figure \ref{fig:plot_age_color_avg_slope} we compare them with the \BV~colors of the sample from \citet{Taylor:2005} that consists of 142 nearby galaxies with ground-based UV and optical imaging. To do so, we calculate the \BV~colors (listed in Table \ref{tab:sersic}) using synthetic \Vb-band magnitudes derived from the SED models. From Figure \ref{fig:plot_age_color_avg_slope} we see that the mean \BVgrad~value of our \mergsz~merger systems is offset toward negative color gradients from that of the Elliptical+Lenticular sample (by \stddeltaBVearlyoff$\sigma$) as expected based on both their merger nature and the average Sersic indices that suggest late- and intermediate type morphologies. While the Spiral+Irregular+Merger sample has a similar mean \BVgrad~value as the Elliptical+Lenticular sample, it has a much larger scatter and hence our merger systems are more consistent with it. However, they still fall on the negative end of the distribution (offset by \stddeltaBVlateoff$\sigma$), showing that our merger systems have relatively negative color gradients compared to typical Spiral+Irregular+Merger galaxies. Therefore, the merger-induced \SF~is preferentially occurring at large radii in our \mergsz~systems.

We also acknowledge the possibility that light from star formation at bluer wavelengths is affected by the presence of dust that is not accounted for in our SED models. As a result, the color gradients may be affected by the presence of dust at small radii that obscures some of the nuclear \SF. Indeed, circumstantial evidence for this effect includes dust lanes seen in some of the merger systems (Figure \ref{fig:HST_color}). In this case, the true effect of merger-induced nuclear \SF~is higher than that inferred from Figure \ref{fig:plot_age_color_avg_slope}. However, Figure \ref{fig:plot_lrt_sdss} shows that the SFRs and stellar masses from our SED modeling are not systematically under-estimated relative to the dust-corrected values from the SDSS spectra. Additionally, the optical selection of AGN and \SF~is known to preferentially target galaxies with relatively small amounts of nuclear gas and dust compared to selections at X-ray or IR wavelengths \citep{Ellison:2016} so that these systems are unlikely to be impacted by significant nuclear obscuration. Finally, Figure \ref{fig:plot_age_color_avg_slope} shows that the mean value of \BVavg~for our merger systems falls on the blue end of the distribution for the Spiral+Irregular+Merger sample (offset by \stdavgBVlateoff$\sigma$). Even after removing the two \typeI~AGN (the two systems with the bluest nuclear colors), the mean \BVavg~value of our merger sample (\meanavgBVreddest) is bluer than the Spiral+Irregular+Merger sample (\meanavgBVlate) and thus does not show an excess of red color at small radii steepening the color gradients. Therefore, we ultimately consider the effect of dust reddening to have a negligible impact on our measured negative color gradients and \SFR s.

\subsection{Connection between \SF~and Morphology}
\label{sec:correlations}

If galaxy interactions and mergers contribute to \SF~and hence affect the average stellar population ages (i.e. through tidally induced torques on gas), then enhancements in global \sSFR~may be correlated with morphological properties of the merger. To test this prediction, we compute the differences between \sSFR~in the merger-selected systems of this work and those of the matched control sample: \sSFRlrtnorm$=$\sSFRlrtnormdef. We have chosen to investigate the morphological parameters of mass ratio (\mratio, from \paperI) and residual asymmetry (\Arot, from Section \ref{sec:galfit_model}).

To test for correlations between \sSFRlrt~and each of the above parameters \mratio~and \Arot, we determined the best-fit linear functions for each set of values. Confidence intervals on the slope and intercept are measured by sampling random errors for each data point from a simulated Gaussian distribution with sigma equal to the true error. We iteratively increase the number of simulations until the lower and upper errors, determined from the $34\%$ lower and upper quantiles, respectively, changed by $<10\%$. While we find that the linear slope between \sSFRlrtnorm~and \mratio~is consistent with a negative correlation, the significance level is $<2\sigma$. On the other hand, the linear slope between \sSFRlrtnorm~and \Arot~is stronger and consistent with a positive correlation at a significance level of \sSFRnormresidfracsig~(top panel of Figure \ref{fig:plot_sSFR_residfrac}).

Since the contamination from scattered light, and hence over-estimates of \sSFRlrt, are larger among the \mergsz~merger systems than the control samples (Section \ref{sec:build_control}), the values of \sSFRlrtnorm~are likely also over-estimated. To consider this effect, we have corrected each value of \sSFRlrtnorm~by the differences in \loiii~luminosity (assuming that the \sSFRlrt~over-estimates scale directly with the scattered light contribution and hence the AGN luminosity). This correction results in qualitatively similar correlations where the linear slope of \sSFRlrtnorm~with \mratio~is consistent with a negative correlation at a significance level of $<2\sigma$. As with the uncorrected values, the linear slope between \sSFRlrtnorm~and \Arot~is the strongest, consistent with a positive correlation at a significance level of \sSFRnormresidfracsigCORR. This is shown in the bottom panel of Figure \ref{fig:plot_sSFR_residfrac}. While the only conclusive correlation among either the uncorrected or corrected \sSFRlrtnorm~values is found with \Arot, both morphological parameters (\mratio~and \Arot) evolve with \sSFRlrtnorm~such that more violent mergers correspond to enhanced \SF.

\subsection{Comparison with the Control Samples}
\label{sec:control}

\begin{figure}[t!]
\hspace*{-0.1in} \includegraphics[width=3.5in]{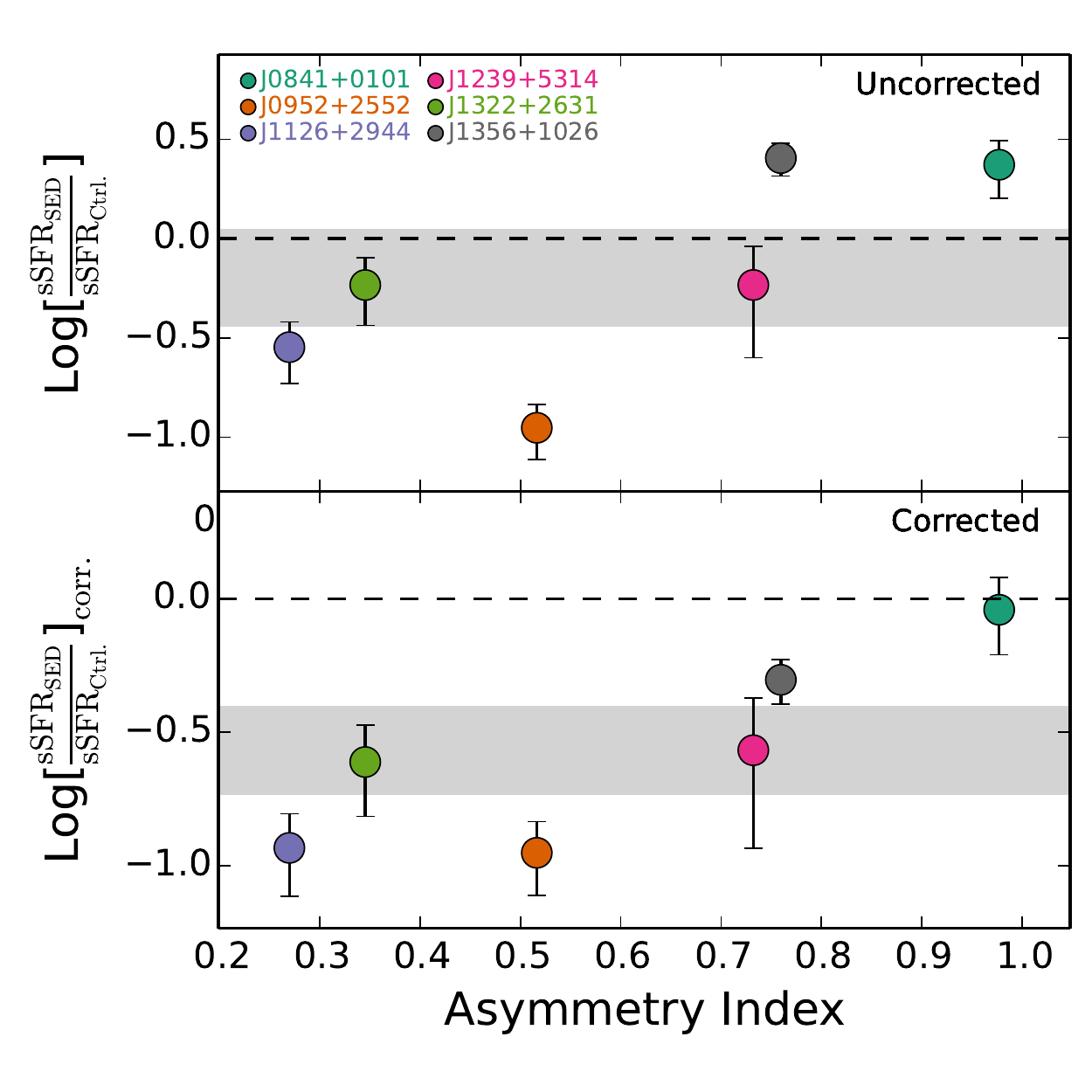}
\vspace*{-0.25in}
\caption{\footnotesize{\sSFRlrtnorm~plotted against \Arot~for the original values (top) and the values corrected for AGN scattering (bottom). The vertical error bars denote the $1\sigma$ confidence intervals based on the standard deviation of the control sample. The dashed line represents \sSFRlrt$=$\sSFRctrl~while the gray shaded region represents the upper and lower $1\sigma$ bounds. The uncorrected and corrected values of \sSFRlrtnorm~increase with \Arot~at significance levels of \sSFRnormresidfracsig~and \sSFRnormresidfracsigCORR, respectively.}}
\label{fig:plot_sSFR_residfrac}
\end{figure}

From Figure \ref{fig:plot_sSFR_residfrac} the average value of \sSFRlrtnorm~is \sSFRlrtavg~with an offset below zero significant at \sSFRlrtavgoff. For comparison, using the corrected values of \sSFRlrtnorm~results in an average of \sSFRlrtavgCORR~with an offset below zero significant at \sSFRlrtavgCORRoff. These results show that, on average, these merger systems are experiencing comparable (uncorrected values) or lower (corrected values) levels of \SF~compared to galaxies with similar masses, redshifts, and AGN luminosities that were not selected to be in merger systems. 

\begin{figure}[t!]
\hspace*{-0.1in} \includegraphics[width=3.5in]{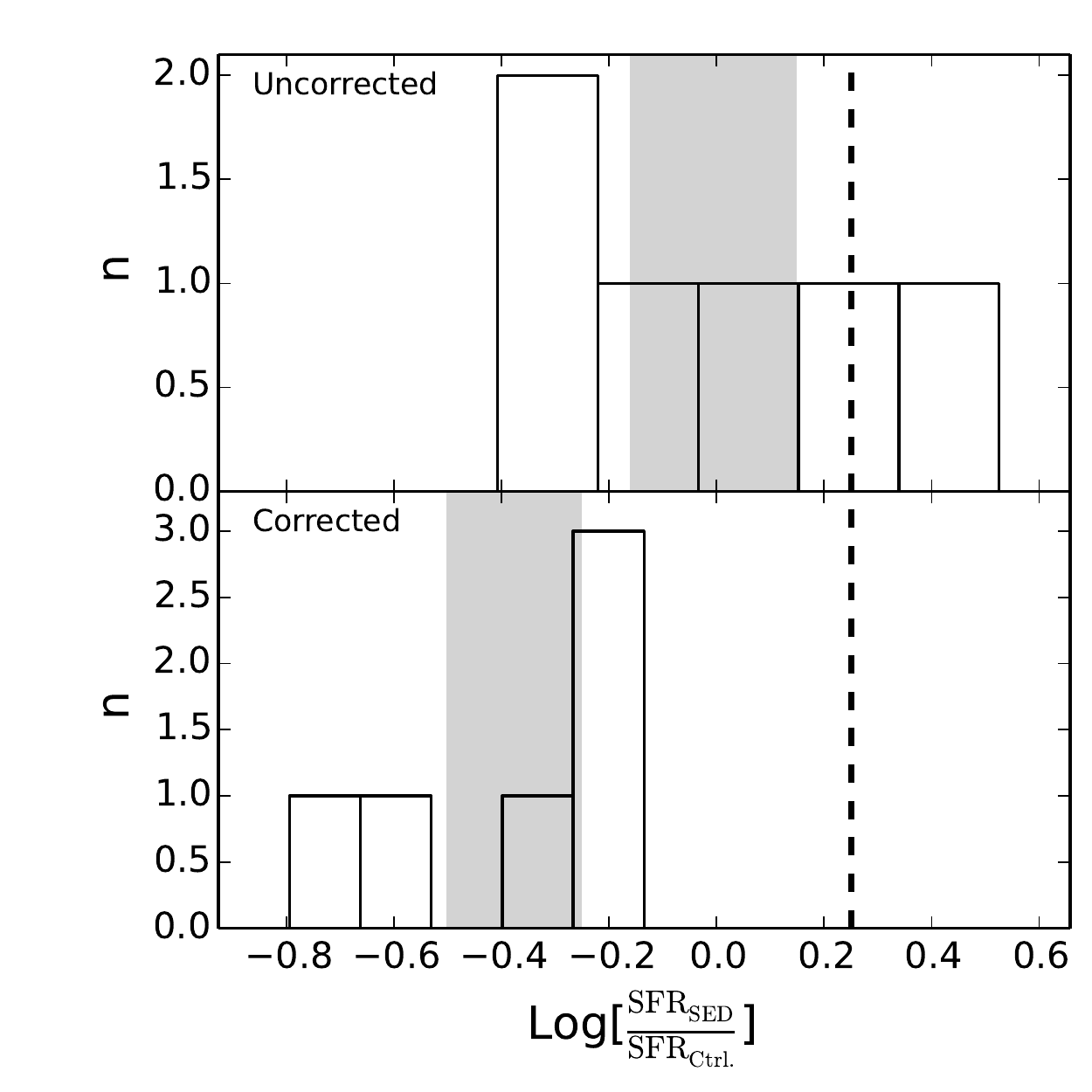}
\vspace*{-0.25in}
\caption{\footnotesize{Distributions of \sSFRlrtnorm~for the uncorrected (top) and corrected (bottom) values. The gray-shaded region represents the standard deviation upper and lower bounds. The value from the smallest separation in Figure 3 of \citet{Scudder:2012} is shown by the vertical black, dashed line ($0.25$). Note that the mean value of \sSFRlrtnorm~is offset below that value from \citet{Scudder:2012} by \SFRlrtavgoffscudder~(uncorrected) and \SFRlrtavgCORRoffscudder} (corrected).}
\label{fig:hist_SFR_UNCORR_CORR}
\end{figure}

To understand these effects on \SF~within the context of the overall galaxy merger process, we compare the normalized values of \SFRlrt~(\SFRlrtnorm$=$\SFRlrtnormdef) to those of earlier-stage merger systems. This is shown in Figure \ref{fig:hist_SFR_UNCORR_CORR} where we compare \SFRlrtnorm~for our sample to that of the smallest separation bin in Figure 3 of \citet{Scudder:2012}, where values of \SFR~are based on the SDSS fiber spectroscopy of galaxies with optical emission lines dominated by \SF. This bin contains galaxies at a larger average separation of $\sim10$ kpc compared to our sample ($4.7$ kpc). We see that the mean values of our sample are offset below that of \citet{Scudder:2012} at significance levels of \SFRlrtavgoffscudder~(uncorrected values) and \SFRlrtavgCORRoffscudder~(corrected values). Since the average separation of our sample is smaller than that of \citet{Scudder:2012}, this result suggests that selection for these systems to be in late-stage galaxy mergers is responsible for marginally (uncorrected values) or significantly (corrected values) smaller \SFR~enhancements relative to slightly earlier-stage mergers. While our sample was selected based upon AGN that dominate the optical emission line spectrum (as opposed to \SF~dominated optical emission lines in the star-forming sample), the smaller \SFR~enhancements are unlikely to be due to this selection effect since we have normalized by control samples of AGN selected in the same way.

We note that low-gas fractions among the merger systems would also result in suppressed \SF. Indeed, our sample consists of optically selected SDSS AGN that likely have low gas supplies, and hence suppressed \SFR, compared to star-forming galaxies \citep{Ellison:2016}. However, this selection effect does not impact our analysis since we have normalized the \SFR s by control samples of AGN that are also selected optically. Additionally, while \citet{Shangguan:2016} suggest that selection of X-ray AGN in galaxy mergers may preferentially find gas-poor systems, this effect is not relevant to our sample since the original AGN identifications were based on optical emission lines. Finally, since the control samples were matched on redshift and mass, both of which strongly affect galaxy gas fractions \citep{Dave:2011}, our results are likely unaffected by evolution of the gas fraction.

For the possibility discussed in Section \ref{sec:radii} where dust is obscuring significant nuclear \SF, the true \SFR~enhancements will be higher than shown in Figure \ref{fig:hist_SFR_UNCORR_CORR} (assuming the control samples are not affected by the same level of nuclear dust obscuration. In this scenario, the SFR enhancements of our late-stage mergers will be more consistent with the larger separation pairs from \citet{Scudder:2012}. However, as described in Section \ref{sec:radii}, the possibility of significant nuclear dust and hidden \SF~is unlikely. Furthermore, we see the same effect when using the subset of four systems with \SFR s from SDSS optical spectra. Since this is the same measurement procedure for \SFR s as in \citet{Scudder:2012}, both samples will be affected by comparable levels of dust \citep{Ellison:2016}.

\section{Discussion}
\label{sec:discussion}

In this section we synthesize the results from Section \ref{sec:results} to understand how the stellar populations in these systems are affected by the mergers and if their evolution is correlated with that of the SMBHs. Specifically, in Section \ref{sec:correlated} we discuss evidence for merger-induced \SF, in Section \ref{sec:time} we discuss how the how \SF~may be declining, and in Section \ref{sec:AGN_SF_connection} we discuss how the merger-induced \SF~and AGN are correlated but may also be offset in time.

\subsection{Evidence of Merger-induced \SF}
\label{sec:correlated}

The \Hb-band imaging analysis in Section \ref{sec:galfit_model} provides evidence suggesting the presence of merger-induced morphological disturbances. First, $4/6$ systems are classified as major mergers, and $5/6$ with mass ratios $<7:1$ \citep{comerford:2015}. Major galaxy mergers are theoretically predicted and observationally shown to result in significant morphological disturbances. Second, the Sersic indices for our \mergsz~systems (Table \ref{tab:sersic}) span the range $n\sim1-4$, and two-thirds of them have Sersic indices in the range \n~$\sim1-2.5$, similar to the late-stage mergers from \citet{Shangguan:2016} and typical of late- and intermediate type galaxies. Intermediate-type morphologies are representative of transitional phases often seen in mergers (Fan et al. 2016). There is also only one case of disk structure (\galaxythree), and it is a very minor merger. Hence any disk structure that existed before the mergers was likely destroyed, suggesting that the morphologies are affected by the on-going mergers. These morphological disturbances may be connected to the evidence of on-going global \SF~presented in Section \ref{sec:radii}. This connection is quantified in Section \ref{sec:correlations} where the significant positive correlation between enhanced \sSFR~and residual asymmetry suggests that the \SF~is dependent on the level of tidal disturbances.

While we also see marginal evidence for enhanced \sSFR s~in more major mergers, the correlation is statistically weak ($<2\sigma$) and hence we do not claim that it is real. The absence of such a correlation would be consistent with previous observations that have found minor mergers to play a role comparable to or greater than that of major mergers in triggering \SF~\citep{Woods:2007,Shabala:2012,Kaviraj:2014,Willett:2015}. Instead, the morphological asymmetries may be more direct tracers of the gravitational tidal forces on the host galaxy potential, a parameter that has been shown to affect \SF~in mergers \citep{Woods:2007}. Hence, they are imprinted with the past dynamics of the merger and lead to the strong correlation with \sSFR~enhancement seen in Figure \ref{fig:plot_sSFR_residfrac}. This connection between mergers and \SF~is consistent with results for earlier stage pairs \citep{Ellison:2008,Scudder:2012,Patton:2013}, and our sample shows that this connection persists at separations of $\sim2$ kpc.

\subsection{Merger-induced Global \SF~May Be Declining}
\label{sec:time}

In Section \ref{sec:control} we showed that the average \sSFR s of our AGN sample are comparable to or lower than those of the control sample (Figure \ref{fig:plot_sSFR_residfrac}) and that the average \SFR~enhancements are lower than those of earlier-stage mergers of star-forming galaxies by $1-5\sigma$ (Figure \ref{fig:hist_SFR_UNCORR_CORR}). Furthermore, we showed that the relative lack of \SFR~enhancements are not due to dust obscuration hiding nuclear \SF~(Section \ref{sec:radii}) or due to a selection bias of AGN dominated optical emission lines (Section \ref{sec:control}).

Therefore, compared to the star-forming samples, the \SF~in our sample may correspond to starbursts induced by the galaxy interaction at a relatively earlier stage and that occurred on larger physical scales \citep{Patton:2013}. Indeed, our analysis in Section \ref{sec:radii} shows that the merger-induced \SF~is preferentially occurring at large radii. On the other hand, the \SFR s used in the star-forming samples are known to be centrally concentrated \citep{Patton:2011}. Therefore, we are likely viewing these late-stage mergers as the global \SFR s are declining relative to the earlier-stage systems. This scenario is predicted by numerical simulations of mergers that find a peak in global \SFR~occurs a few Myrs after the first pericentric passage when the nuclear separation ranges from $10-100$ kpc \citep{Cox:2008,Kim:2009,Teyssier:2010,Scudder:2012,Stickley:2014,Renaud:2014}. As the merger progresses to nuclear separations below $10$ kpc (corresponding to the separations in our sample), the global \SF~has mostly completed \citep{Patton:2011} and is predicted to be relatively unaffected compared to the nuclear \SFR~\citep{Hopkins2008,Capelo:2015}.

The nuclear \SFR~may continue to rise toward later merger stages (potentially peaking post-coalescence) when the AGN also becomes more active, leading to correlated activity \citep{Li:2008}. Indeed, a subsequent rise in nuclear \SF~that corresponds temporally to the AGN peak luminosity is numerically predicted \citep{Hopkins:2012a}, and simulations  show that this nuclear \SF~can account for a significant fraction of the overall galaxy \SFR~\citep{Capelo:2015,Volonteri:2015}. Therefore, the nuclear \SFR~enhancements in our sample may continue rising toward later merger stages and eventually peak after nuclear coalescence.

\begin{figure}[t!]
\hspace*{-0.1in} \includegraphics[width=3.5in]{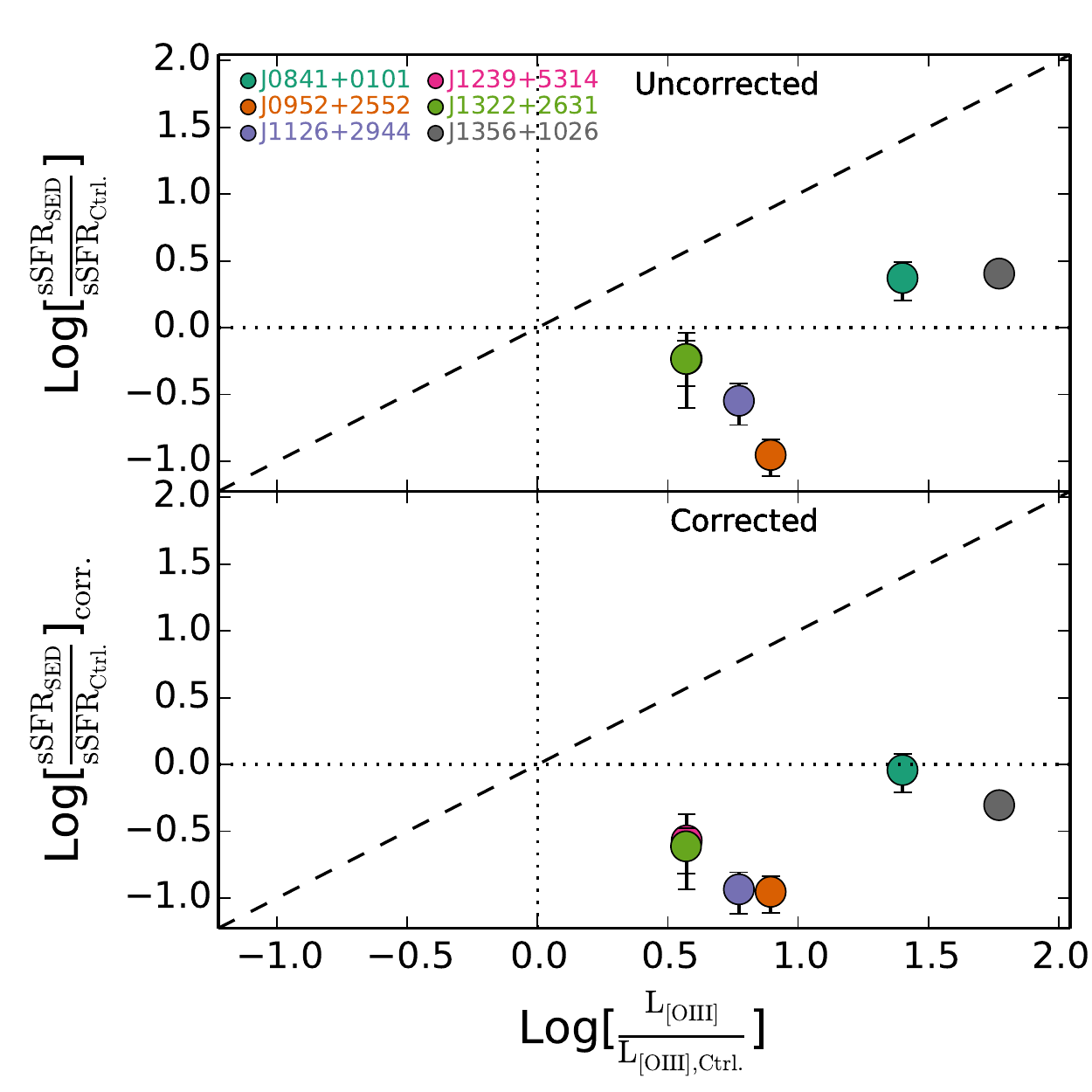}
\caption{\footnotesize{\sSFRlrtnorm~plotted against \LOIIInorm~for the uncorrected (top) and corrected (bottom) values. The dashed line represents the one-to-one relation. The horizontal and vertical dotted lines represent \sSFRlrt$=$\sSFRctrl~and \loiii$=$\loiiictrl, respectively. The uncorrected and corrected values of \sSFRlrtnorm~increase with \LOIIInorm~at significance levels of \sSFRnormLOIIInormsig~and \sSFRnormLOIIInormsigCORR, respectively.}}
\label{fig:plot_sSFR_LOIII_UNCORR_CORR}
\end{figure} 

We note that the offset in Figure \ref{fig:hist_SFR_UNCORR_CORR} of \SFRlrtavgoffscudder~does not robustly eliminate the possibility that the normalized \SFR s for our sample are in fact similar to those of the larger separation pairs. However, we can still confidently rule out the possibility that the normalized \SFR s in our sub~$-10$ kpc separation sample are elevated relative to larger separation pairs. Thus when viewed together, the studies are consistent with a scenario in which the enhancements in global \SFR~subside below $\sim10$ kpc.

\subsection{Connection between \SF~and AGN in Late-stage Galaxy Mergers}
\label{sec:AGN_SF_connection}

In this section we investigate the connection between \SF~and AGN triggering in our \mergsz~late-stage merger systems. Specifically, we examine the connection between merger-induced \SF~(Section \ref{sec:AGN_SF_correlation}) but also discuss possible evidence for a delay between the two (Section \ref{sec:delay}).

\subsubsection{Both \SF~and AGN Are Merger-triggered}
\label{sec:AGN_SF_correlation} 

The discussion in Section \ref{sec:correlated} shows that enhancements in \sSFR~are strongest in more disturbed systems, a result strongly predicted by observational results from merger samples at earlier stages \citep{Ellison:2008,Patton:2011,Scudder:2012}. Furthermore, some observational evidence suggests that these same violent mergers are more likely to trigger AGN as well \citep{Treister:2012,Glikman:2015}. From these two observations follows the prediction that merger-induced \SF~and AGN scale with each other. However, as mentioned in Section \ref{sec:intro}, finding observational evidence of this correlation has proven difficult. Numerical simulations from \citet{Volonteri:2015} predict that the link between global \SFR~and SMBH accretion rate only emerges among their late-stage merger sample. Indeed, our sample contains global \SFR s that are consistent with the \SFR s of the \citet{Volonteri:2015} late-stage merger sample ($1-10$ \Mu~yr$^{-1}$) and hence is suitable for testing this prediction. 

\begin{figure}
\hspace*{-0.1in} \includegraphics[width=3.5in]{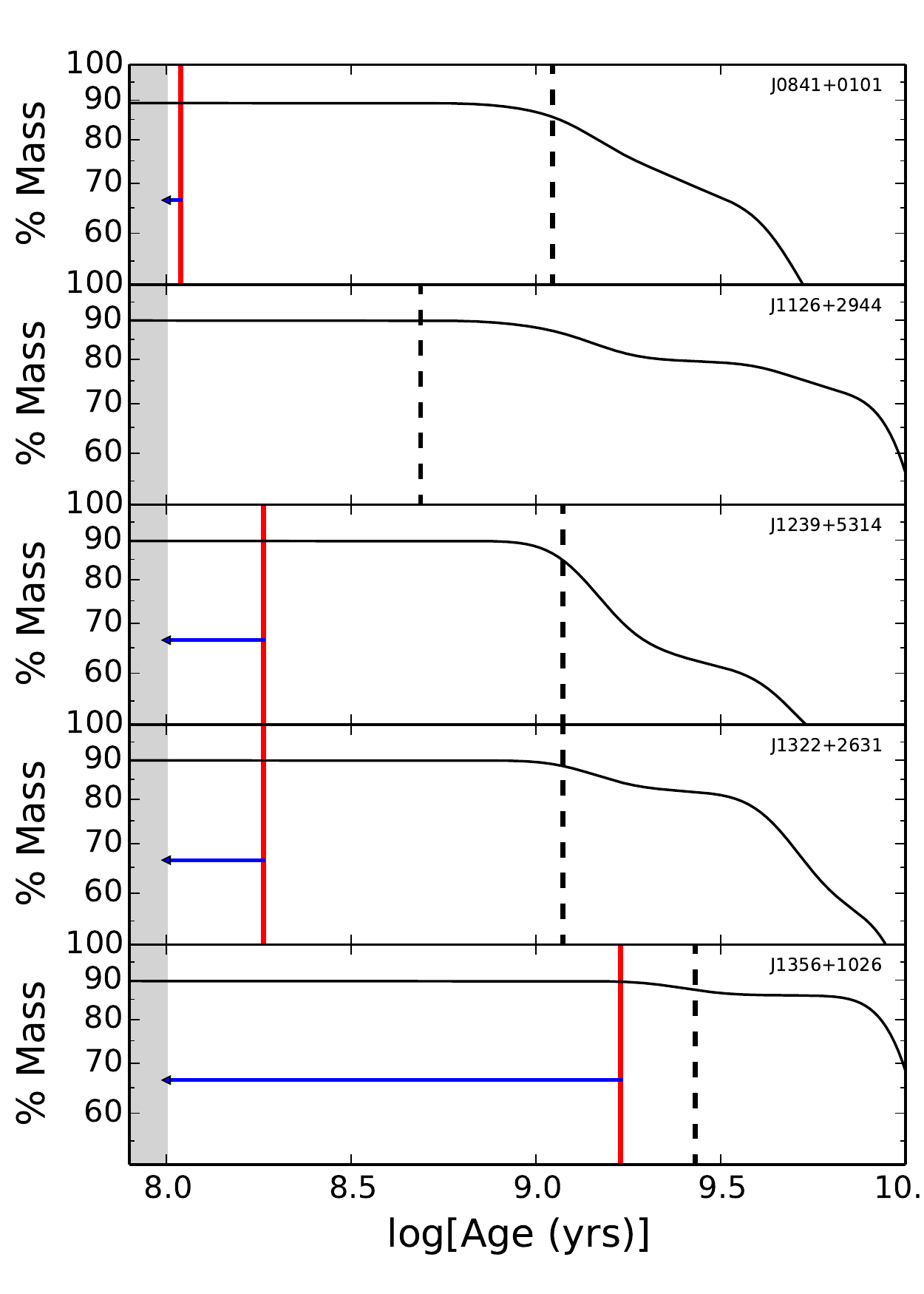}
\caption{\footnotesize{Delays between the peak of merger triggered \SF~and the onset of merger triggered AGN for the five galaxies with reliable \strlt~models. The solid black curve represents the fraction of mass assembled from star formation as a function of lookback time based on the ages of stellar bases in the best-fit \strlt~models The vertical dashed, black line represents the estimated onset of the merger, the vertical solid, red line represents the estimate peak of global \SF, and the gray shaded region represents the estimated AGN duty cycle. The horizontal blue arrow represents the temporal delay between the estimated peak of global \SF~and the onset of the AGN duty cycle in cases where a delay is detected. All times and ages are in the galaxy's reference frame.}}
\label{fig:plot_base_age_Mcor}
\end{figure}

To do so, we have created control samples for each of the AGN in a manner identical to that of Section \ref{sec:build_control} except that we have excluded the \loiii~matching criteria so that we can compare the AGN bolometric luminosities (assumed to be proportional to \loiii). We then compute the normalized values of \loiii~(\LOIIInorm$=$\LOIIInormdef) and plot \LOIIInorm~against \sSFRlrtnorm~in Figure \ref{fig:plot_sSFR_LOIII_UNCORR_CORR}. We see that all values of normalized \loiii~are significantly above unity, a result that is likely due to the selection of bright AGN for \ch~detections (see Section \ref{sec:sample}). We test for a correlation between normalized \loiii~and normalized \sSFR~using the same procedure as in Section \ref{sec:correlations}, finding positive correlations at significance levels of \sSFRnormLOIIInormsig~and \sSFRnormLOIIInormsigCORR~for the uncorrected and corrected samples, respectively (we note that the correlation is driven primarily by the two galaxies with the brightest AGN). Therefore, we see evidence that enhancements in \sSFR~correspond to enhancements in AGN luminosity. 

\subsubsection{Time Delay between Merger-induced \SF~and AGN}
\label{sec:delay}

While Figure \ref{fig:plot_sSFR_LOIII_UNCORR_CORR} shows that the normalized values of \loiii~are all significantly above unity, the average value for the normalized \sSFR s is less than unity (discussed in Section \ref{sec:control}). This significant difference in the relative enhancements is consistent with the implication from Section \ref{sec:time} that the merger-induced enhancement in \sSFR~happened at an earlier phase in the merger than for the AGN. Moreover, this result is qualitatively consistent with the detection of delays between \SF~and AGN in observational work \citep{Schawinski:2009,Wild:2010,Kaviraj:2015a,Shabala:2017} and theoretical work \citep{Hopkins:2012a,Capelo:2015}. 

To directly compare our inferred delays with these works, we create a timeline of stellar mass assembly using the best-fitting combination of \strlt~stellar templates following the procedure described in \citet{Fernandes:2007}. For each of the stellar population bases used by our \strlt~modeling the implicit assumption exists that all of the mass was created in a instantaneous burst of \SF~\citep{Fernandes:2005}. Therefore, for a given look-back time, the total stellar mass formed through \SF~is the cumulative sum of the masses for all stellar populations with ages equal to or less than that time. Examining this cumulative sum as a function of stellar population age shows the timescale (in the galaxy's reference frame) over which stellar mass is assembled due to \SF. These timelines are shown in Figure \ref{fig:plot_base_age_Mcor} for each of the \delaysz~galaxies with reliable \strlt~models, where we have normalized the cumulative mass by the total mass from all stellar bases so that it represents the mass fraction.

To place the peak in global \SF~on the timeline we assume that it occurs $10^{9}$ yrs after the merger begins (this value corresponds to an upper estimate based on the range of values from simulations ($300-700$ Myrs) and is typically a few hundred Myrs after the first pericentric passage \citep{Scudder:2012,Stickley:2014,Capelo:2015}. The merger onset corresponds to the time when the fraction of remaining mass to be assembled equals the fraction of mass assembled by merger-induced \SF~(\fMmerg). We estimate a conservatively low value of \fMmerg~from the `burst efficiency' function \citep{Cox:2008}, the merger mass ratios, and a low estimate of $10\%$ for the galaxy gas mass fractions \citep{Jaskot:2015}. The lookback times of the merger onset are shown in Figure \ref{fig:plot_base_age_Mcor}, and they represent conservatively low estimates.

Since the AGN are observed to currently be active, the maximum lookback time for their merger-induced onset is $10^{8}$ \ageu s based on typical estimates of AGN duty cycles \citep{Parma:2007,Shulevski:2015}. Figure \ref{fig:plot_base_age_Mcor} shows that four of the five systems with reliable \strlt~models (\galaxyone, \galaxyfour, \galaxyfive, and \galaxysix) have lower limits on the peak global \SF~lookback times (log[\age~(yrs)]~$=8.04-9.23$) that are greater than the AGN duty cycle of $10^{8}$ yrs. These times correspond to lower estimates of the delay between the global \SF~peak and AGN onset of log[\age~(yrs)]~$=6.96-9.20$, with an average value of $4.4\times 10^{8}$ yrs.

We note that the average value of $4.4\times 10^{8}$ yrs is similar to the delays of $100-400$ Myrs measured by previous works \citep{Schawinski:2009,Wild:2010,Kaviraj:2015a,Shabala:2017}. However, the average delay of our sample represents a lower estimate and hence the true value is likely to be longer. We hypothesize that our measured time delays are longer because they exclude post-merger systems for which subsequent bursts of small-scale \SF~may occur after nuclear coalescence.

\section{Conclusions}
\label{sec:conclusions}

We have analyzed the \hst~imaging and archival photometry and spectra for a sample of \mergsz~late-stage galaxy mergers hosting AGN. In \paperI, we put constraints on the efficiency of the mergers for triggering the AGN. In this paper, our aim is to understand the effects of the mergers on evolution of the host galaxy stellar populations and if it is correlated with SMBH growth. We have used \hst~imaging to quantify the tidal disturbances in the stars and the spatial distribution of star formation. We have used the photometric and spectral data to measure the star formation rates, stellar masses, and stellar ages of the systems. Our conclusions are as follows:

\begin{enumerate}

\item With increasing radial distance from the nuclei the merger systems become increasingly dominated by younger stellar populations. Furthermore, the radial color gradients are on the negative end of the distribution for typical galaxy mergers suggesting that the majority of merger-induced star formation is occurring on a galaxy-wide (global) scale.

\item The specific star formation rates, normalized by matched control samples, are positively correlated with asymmetries in the NIR images. This result suggests that enhancements in the specific star formation rates are strongly coupled to the level of tidal disturbances. Hence, the elevated specific star formation rates are likely to be merger-induced.

\item The normalized star formation rates are on average lower than those from larger separation (earlier merger stage) galaxy pairs by $>1\sigma$ and potentially by $\sim5\sigma$ when estimates of scattered AGN flux are taken into account. An offset toward smaller star formation rate enhancements compared to larger separation pairs is consistent with galaxy merger simulations predicting a decline in global star formation rates below separations of $\sim10$ kpc.

\item Enhancements in specific star formation rates are positively correlated with enhanced AGN luminosity, suggesting that both values are mutually triggered by the merger events. This result is consistent with predictions from late-stage merger simulations and with extrapolations from previous studies of larger separation pairs that imply AGN luminosity and star formation are both enhanced in galaxy mergers.

\item The average enhancement in AGN luminosity is significantly larger than that of the specific star formation rates, suggesting that the level of AGN triggering in these late-stage systems exceeds that of star formation. Furthermore, in four out of five systems we see evidence for an average difference of $\gtrsim 10^{8}$ yrs between the peak of global, merger-induced star formation and the onset of SMBH accretion. Both results are consistent with delays between merger-triggered star formation and AGN.

\end{enumerate}

Support for this work was provided by NASA through Chandra Award Number GO2-13130A issued by the Chandra X-ray Observatory Center, which is operated by the Smithsonian Astrophysical Observatory for and on behalf of NASA under contract NAS8-03060. Support for HST program number GO-12754 was provided by NASA through a grant from the Space Telescope Science Institute, which is operated by the Association of Universities for Research in Astronomy, Inc., under NASA contract NAS5-26555.
The scientific results reported in this article are based in part on observations made by the Chandra X-ray Observatory, and this research has made use of software provided by the Chandra X-ray Center in the application packages CIAO, ChIPS, and Sherpa. The results reported here are also based on observations made with the NASA/ESA Hubble Space Telescope, obtained at the Space Telescope Science Institute, which is operated by the Association of Universities for Research in Astronomy, Inc., under NASA contract NAS 5-26555. These observations are associated with program number GO-12754.

\end{document}